\documentclass[5p,authoryear]{elsarticle}

\usepackage{bm}
\usepackage{amsmath}
\usepackage{subfigure}
\usepackage{hyperref}
\hypersetup{
  colorlinks = true,
  citecolor = blue,
  linkcolor = blue 
}
\usepackage[hyperref]{xcolor}
\usepackage{xstring}
\usepackage{makecell}

\begin{document}

\title{RadioAstron orbit determination and evaluation of its results using
  correlation of space-VLBI observations}

% TBD
\author[kiam]{M.~V.~Zakhvatkin\corref{mycorrespondingauthor}}
\cortext[mycorrespondingauthor]{Corresponding author at: Keldysh Institute of
  Applied Mathematics, Russian Academy of Sciences, Miusskaya sq. 4, 125047
  Moscow, Russia.}
\ead{zakhvatkin@kiam1.rssi.ru}
\author[asc]{A.~S.~Andrianov}
\author[asc]{V.~Yu.~Avdeev}
\author[asc]{V.~I.~Kostenko}
\author[asc]{Y.~Y.~Kovalev}
\author[asc]{S.~F.~Likhachev}
\author[asc]{I.~D.~Litovchenko}
\author[gaish,asc,bstu]{D.~A.~Litvinov}
\author[asc]{A.~G.~Rudnitskiy}
\author[asc]{M.~A.~Shchurov}
\author[asc]{K.~V.~Sokolovsky}
\author[kiam]{V.~A.~Stepanyants}
\author[kiam]{A.~G.~Tuchin}
\author[asc]{P.~A.~Voitsik}
\author[kiam]{G.~S.~Zaslavskiy}
\author[gaish,phys_msu]{V.~E.~Zharov}
\author[asc]{V.~A.~Zuga}

\address[kiam]{Keldysh Institute of Applied Mathematics RAS, Miusskaya sq. 4,
  Moscow, Russia 125047}
\address[asc]{Astro Space Center of Lebedev Physical Institute of RAS, 84/32
  Profsoyuznaya st., Moscow, Russia 117997}
\address[gaish]{Sternberg Astronomical Institute, Lomonosov Moscow State
  University, Universitetsky pr. 13, Moscow, Russia 119991}
\address[phys_msu]{Faculty of Physics, Lomonosov Moscow State University,
  Leninskie Gory, Moscow, Russia 119991}
\address[bstu]{Bauman Moscow State Technical University, 2-ya Baumanskaya 5,
  Moscow, Russia 105005}

\begin{abstract}
  A crucial part of a space mission for very-long baseline interferometery
  (VLBI), which is the technique capable of providing the highest resolution
  images in astronomy, is orbit determination of the mission’s space radio
  telescope(s). In order to successfully detect interference fringes that
  result from correlation of the signals recorded by a ground-based and a
  space-borne radio telescope, the propagation delays experienced in the
  near-Earth space by radio waves emitted by the source and the relativity
  effects on each telescope’s clock need to be evaluated, which requires
  accurate knowledge of position and velocity of the space radio telescope. In
  this paper we describe our approach to orbit determination (OD) of the
  RadioAstron spacecraft of the RadioAstron space-VLBI mission. Determining
  RadioAstron’s orbit is complicated due to several factors: strong solar
  radiation pressure, a highly eccentric orbit, and frequent orbit
  perturbations caused by the attitude control system. We show that in order
  to maintain the OD accuracy required for processing space-VLBI observations
  at cm-wavelengths it is required to take into account the additional data on
  thruster firings, reaction wheel rotation rates, and attitude of the
  spacecraft.We also investigate into using the unique orbit data available
  only for a space-VLBI spacecraft, i.e. the residual delays and delay rates
  that result from VLBI data processing, as a means to evaluate the achieved
  OD accuracy. We present the results of the first experience of OD accuracy
  evaluation of this kind, using more than 5,000 residual values obtained as a
  result of space-VLBI observations performed over 7 years of the RadioAstron
  mission operations.
\end{abstract}


\maketitle{}

\section{Introduction}
\label{intro}
The RadioAstron spacecraft, equipped with a 10-m space radio telescope (SRT),
was launched into a highly elliptical Earth orbit in July 2011 by means of a
Zenit-3F rocket and Fregat-SB upper stage. The main purpose of the spacecraft
is to conduct observations of galactic and extragalactic radio sources in
conjunction with ground-based radio telescopes forming a multi-antenna
ground-space radio interferometer with extremely long baselines
\cite{Kardashev2013}. The spacecraft also made it possible to test the
Einstein Equivalence Principle thanks to its eccentric orbit and a highly
stable on-board hydrogen maser frequency standard \cite{litvinov2018pla,
  NUNES2019}.

The SRT observes at the four frequency bands: P-, L-, C- and K. During
observations the RadioAstron spacecraft transmits the science data being
collected in real time to a tracking station on Earth using its 1.5-m
high-gain dish antenna via a 15~GHz downlink. The science payload of the SRT
includes two hydrogen maser frequency standards (H-masers) with only one of
the two allowed to operate at a time. Shortly after launch one of the two
H-masers was identified as impaired, while the other fully functional. The
latter H-maser was thus used since the beginning of the mission operations in
2011 till August 2017, when it exhausted its hydrogen supply and was switched
off, thereby exceeding its expected lifetime by a factor of two. The science
payload of the SRT and its radio downlinks may be synchronized either to the
on-board H-maser signal or to a ground H-maser signal uplinked from one of the
RadioAstron mission's tracking stations. While the on-board H-maser was
operational, its signal, being slightly more stable than that of the uplink,
was used as a reference both for the science payload and the downlink signals,
including those analyzed in this paper.

A C-band radio link is used for telemetry and command as well as to obtain
range and Doppler tracking data for the orbit determination needs. Routine
tracking is performed every 2--3 days by the two antennas located in Russia:
the 64-m antenna at Bear Lakes (near Moscow) and the 70-m antenna at Ussuriysk
(Primorsky Krai). RadioAstron is also equipped with a retroreflector array,
which makes it possible to perform satellite laser ranging on it.

RadioAstron's orbit is highly elliptic, with the geocentric distance varying
from 7,000 to 351,600 km and an average period of 8.6 day.  The orbit of the
spacecraft significantly evolves with time due to the gravitational pull of
third bodies. The orbital parameters change periodically with a period of
about 2 years, which provides for rich opportunities of observing various
radio sources distributed all over the sky.

VLBI data processing involves the so-called fringe search procedure, i.e. the
search for a maximum of the cross-correlation of the signals recorded at
different telescopes. This procedure requires the delay in arrival times of
the wavefront at each pair of telescopes to be accurately known for the
duration of the observation. The initial guess for the delay is calculated
according to a model which utilizes various kinds of data: trajectories of the
phase centers of the participating telescopes in an inertial reference frame,
offsets and drifts of their clocks, propagation media parameters, etc. In
order to be able to perform the fringe search in a reasonable amount of time,
e.g. tens of minutes, using a modern computing cluster, a priori uncertainty
in the parameters of the model should be relatively small and known (the
required level of uncertainties depends on many factors, a few examples are
given below).

In space-VLBI the major uncertainty in the modeled delay and its derivatives
comes from orbit reconstruction errors of the spacecraft(s) carrying the radio
telescope(s). For the RadioAstron spacecraft the following requirements to the
OD accuracy were established prior the launch:
\begin{itemize}
\item position error less than 600~m;
\item velocity error less than 2~cm/s;
\item acceleration error less than 10$^{-8}$~m/s$^2$.
\end{itemize}
These requirements are determined by several factors, including the following:
the wavelength of the observation, the data bit rate, the number of
participating telescopes, the correlator design, the available computational
resources, etc.

The requirement on the position accuracy formally reflects the delay search
window with 128 spectral channels. The actual window size and the number of
channels used by the correlator is significantly larger, e.g. 2048 spectral
channels for the survey of active galactic nuclei (AGN), which allows to
account for different kinds of errors. The velocity error less than 2~cm/s
allows to perform a fringe search on the smallest wavelength (1.35~cm) using
the integration time of only 1/8~s in the correlator. Rather stringent
requirement on the acceleration accuracy initially aroused from the necessity
of keeping the phase error of a K-band signal over integration interval of
$\sim$10 minutes within a fraction of a radian. However the acceleration error
within $\pm1.5\times10^{-6}$ m/s$^2$ window can be compensated by the
correlator during the processing.

For an earlier, and so far the only other, realized space-VLBI mission of
VSOP/HALCA the OD requirements were similar (1-$\sigma$ errors are given):
position --- 80 m, velocity --- 0.43 cm/s, acceleration --- $6\times10^{-8}$
m/s$^2$ \cite{halca_nav}. The similarity is easily understandable as
VSOP/HALCA was designed to observe at comparable wavelengths, 1.35, 6 and 18
cm, and recorded the data at the same bit rate as RadioAstron (128
Mbps). Despite the fact that scientific observations with HALCA were routinely
taken only at 1.6 and 5 GHz because of sensitivity problems at 22 GHz, its OD
requirements were formulated for all three bands. Slightly more stringent
requirements of VSOP to position and velocity determination accuracies are due
to a different correlator design.

Orbit determination of the RadioAstron spacecraft is complicated by limited
tracking support and significant non-gravitational perturbations caused by
solar radiation pressure (SRP) and autonomous firings of thrusters of the
spacecraft attitude control system. A unique feature of a space-VLBI
spacecraft is the possibility of verifying its OD accuracy by using the
so-called residual delays and delay rates obtained as a result of successful
detections of interference fringes and post-correlation analysis. In this work
we summarize the first experience of OD accuracy evaluation based on using
residual data of space-VLBI observations. Contrary to the line-of-sight
tracking observations this kind of data allows to measure errors of the
spacecraft position and velocity projected on different directions. This
analysis allowed us to quantitatively evaluate the accuracy of the two
versions of the OD algorithm used.

\section{Orbit and dynamics}
The RadioAstron spacecraft was launched into a highly elliptic orbit with
apogee distance varying between 280 000 and 351 000 km, orbital period between
8.1 and 10.2 days and perigee height above 650 km. The orbit evolves
significantly with time in both eccentricity and orientation of the orbital
plane because of luni-solar gravitational perturbations and due to the
influence of Earth's gravitational field during occasional low perigees (Table
\ref{tab:orb_elem}). An important aspect of such orbit for space-VLBI is that
it provides for observing a large sample of galactic and extragalactic radio
sources at a wide range of projected baselines. Only three trajectory
correction maneuvers have been performed so far. Their goal was to prolong the
lifetime of the orbit by preventing reentry while passing an upcoming local
minimum of the perigee height and also to prevent unacceptably long shadowing
of the spacecraft.

\begin{table*}[htb]
  \footnotesize \centering
  \caption{Evolution of the orbital elements of the RadioAstron spacecraft.}
  \begin{tabular}{l | r | r | r | r | r | r | r }
    Epoch & 2011/08/21& 2012/10/13& 2014/01/27& 2015/03/23& 2016/06/24& 2017/07/31& 2018/02/17\\
    \hline
    Perigee height, $10^3$ km    & 6.213    & 69.265  & 1.075   & 67.666 & 0.654 & 75.716 & 3.699\\
    Apogee height, $10^3$ km & 336.526  & 278.126 & 338.611 & 280.593 & 336.278 & 306.435 & 329.026\\
    % $e$             & 0.929157 & & & & \\
    $i$, deg        & 56.354   & 76.804  & 9.199   & 56.692  & 43.866 & 68.527 & 51.309\\
    $\Omega$, deg   & 331.682  & 295.970 & 152.397 & 107.007 & 0.930 & 306.140 & 297.661\\
  \end{tabular}
  \label{tab:orb_elem}
\end{table*}

Except for the rare moments when the spacecraft is moving near a perigee and
the perigee height is near its local minimum, the major perturbing
accelerations are due to gravitational pull of third bodies and
non-gravitational accelerations. The latter are the main source of errors when
modeling the dynamics of the RadioAstron spacecraft. Of the non-gravitational
perturbations SRP has the greatest impact on the orbit.

Because of its 10-m on-board antenna, the area-to-mass ratio of the
RadioAstron spacecraft is as high as 0.03 kg/m$^2$. Depending on the
spacecraft orientation with respect to the Sun this results in accelerations
due to SRP as high as 2$\cdot 10^{-7}$ m/s$^2$. Accounting for SRP is
additionally complicated by the fact that the spacecraft has to change its
attitude according to the direction to the radio source to be observed, thus
the SRP may vary significantly over a time span of several hours.

In addition to direct acceleration, SRP produces the major part of the
perturbing torque. Attitude control of the RadioAstron spacecraft, which
includes compensation of external torques, is implemented using reaction
wheels. The spacecraft attitude in the inertial reference frame can be roughly
described as piece-wise constant --- most of the time the spacecraft is not
rotating, with external torques compensated by the rotation of the reaction
wheels, the changes of attitude occur relatively quickly, with their duration
of order of a few minutes.

Several times a day the reaction wheels are desaturated. This happens either
due to the angular momentum accumulated by the reaction wheels reaching its
maximum allowed value, so that they cannot parry the external torque any
further, or due to the need to perform a significant attitude change which
would not be possible using the reaction wheels alone in their current
state. Desaturating the reaction wheels implies stopping them almost
completely with resulting rotation of the spacecraft prevented by means of
thrusters. Every desaturation produces a net Delta V of 3--7 mm/s. Total
effect of the desaturations on the orbit is comparable to the one of direct
SRP and therefore must be taken into account.

\begin{figure}[htb]
  \centering
  \includegraphics[width=\linewidth]{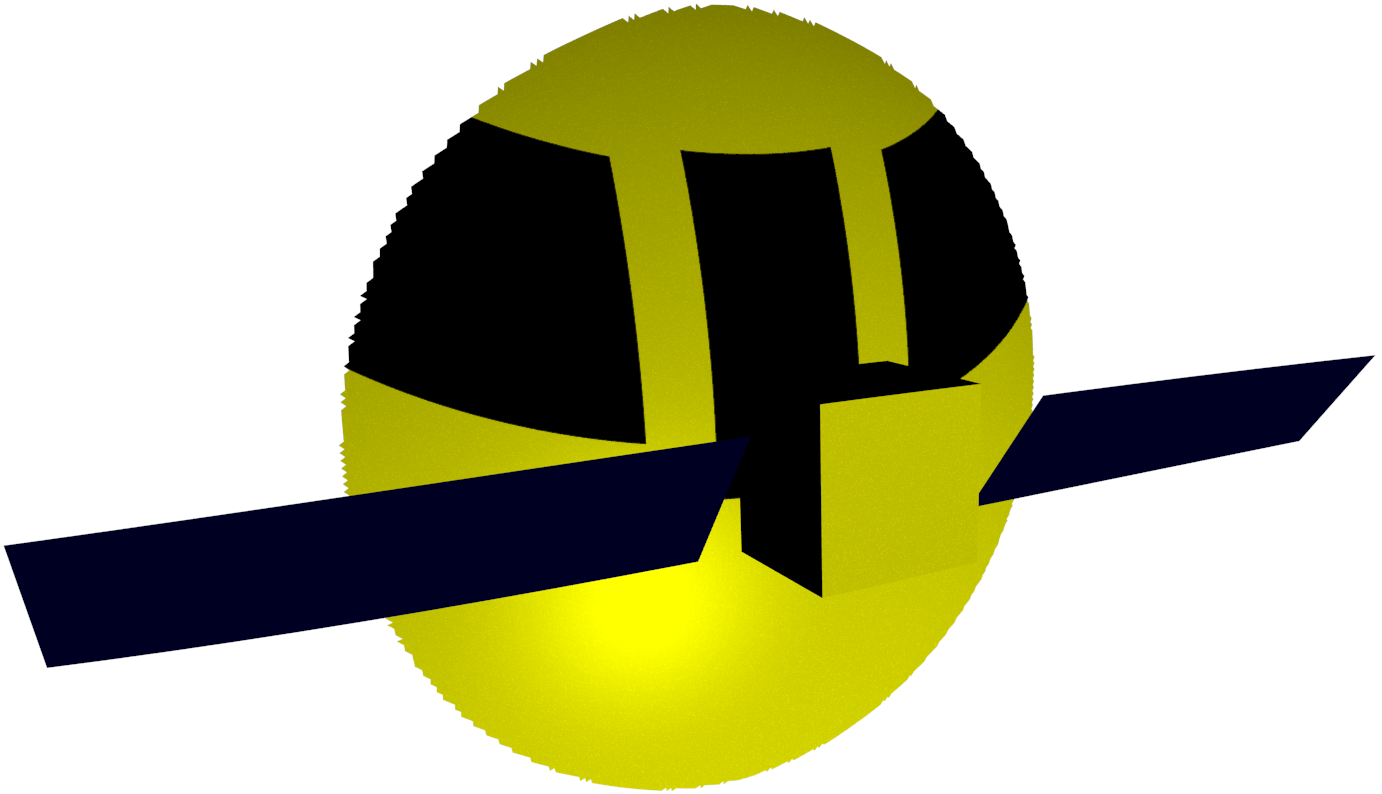}
  \caption{The approximation of the RadioAstron spacecraft surface used in the
    SRP model}
  \label{fig:ra_surf}
\end{figure}

In order to properly take the SRP influence into account, we have developed an
adjustable model, which allows us to calculate both the perturbing
acceleration and torque for a given attitude with respect to the Sun
\cite{zakhvatkin2014SR_RA_eng}. The SRP model utilizes a simplified model of
the spacecraft surface, which consists of the 10-m parabolic dish antenna, the
solar panels and the spacecraft bus (Fig.~\ref{fig:ra_surf}). The coefficients
of reflectivity $\alpha_1$ and specularity $\mu_1$ are associated with sunlit
surfaces of the antenna and spacecraft bus, the coefficient of reflectivity
$\alpha_2$ --- with the surface of the solar panels. According to the model,
the force and torque due to SRP in the spacecraft-fixed frame are
\begin{align}
  \label{eq:srp_f}
  \mathbf{F}_{SRP} = &\alpha_1(1 - \mu_1)\mathbf{F}_{A,B}^{d}(\mathbf{s}) + \alpha_1\mu_1\mathbf{F}_{A,B}^{s}(\mathbf{s}) \nonumber\\
  + &(1-\alpha_1)\mathbf{F}_{A,B}^{a}(\mathbf{s}) \nonumber \\
  + &\alpha_2\mathbf{F}_{SP}^{s}(\mathbf{s}) + (1 - \alpha_2)\mathbf{F}_{SP}^{a}(\mathbf{s}), \\
  \label{eq:srp_t}
  \mathbf{M}_{SRP} = &\alpha_1(1 - \mu_1)\mathbf{M}_{A,B}^{d}(\mathbf{s}) + \alpha_1\mu_1\mathbf{M}_{A,B}^{s}(\mathbf{s}) \nonumber\\
  + &(1-\alpha_1)\mathbf{M}_{A,B}^{a}(\mathbf{s}) \nonumber \\
  + &\alpha_2\mathbf{M}_{SP}^{s}(\mathbf{s}) + (1 - \alpha_2)\mathbf{M}_{SP}^{a}(\mathbf{s}).
\end{align}
Here, $\mathbf{s}$ is the Sun vector, $\mathbf{F}()$ and $\mathbf{M}()$ on the
right-hand side of Equations~\eqref{eq:srp_f} and \eqref{eq:srp_t} represent
the contributions to, respectively, the force and torque from various parts of
the spacecraft, which we denoted by the following subscripts: $A, B$ --- the
antenna and the spacecraft bus, $SP$ --- solar panels. The superscripts $d$,
$s$ and $a$ denote, respectively, diffuse reflection, specular reflection and
absorption of the solar radiation.

The surface of the spacecraft bus and the solar panels in the model consists
of a number of rectangular elements. Provided that all restrictions on the
spacecraft attitude are met, these elements do not shadow each other from the
Sun, which facilitates the calculation of the force and torque components.
Surface of the parabolic antenna of the SRT is represented with 4096 flat
triangular elements. Each element of the antenna surface is considered sunlit
if other parts of the antenna or rectangular elements of the spacecraft bus or
solar panels do not shadow the geometric center of the element. Thus the
surface of the spacecraft in the model consists of flat elements, for each of
which the SRP force due to absorption, specular reflection and diffuse
reflection of radiation is calculated using the area of the element, its
normal vector and unit vector towards the Sun. Summation over all sunlit
elements provides $\mathbf{F}()$ and $\mathbf{M}()$ components in the
right-hand sides of Equations~\eqref{eq:srp_f} and \eqref{eq:srp_t}. These
components depend only on the position of the Sun, $\mathbf{s}$, therefore,
they are tabulated for various Sun angles to speed up the integration of
equations of motion.

Although this model describes the direct SRP perturbation, we believe that
adjustable coefficients allow it to also account for the major part of the
force due to thermal radiation from the spacecraft surface.

We do not use the torque estimation provided by Equation~\eqref{eq:srp_t} in
the model of motion of the spacecraft's center of mass. However, unlike the
SRP force the torque, as will be shown below, can be measured directly. Since
the calculated value of the torque depends on the SRP coefficients,
observations of the torque can be used in the OD to improve the estimate of
the SRP coefficients. The SRP coefficients are assumed constant during the
whole orbit determination interval, which usually spans for 20--30 days.

The complete specification of the dynamic model we use to describe the orbital
motion of the RadioAstron spacecraft is given in
Table~\ref{tab:dyn_model}. Between the events of reaction wheel desaturation
(momentum dumps) the spacecraft moves passively and every desaturation event
is associated with a certain Delta~V vector. A priori values of velocity
change vectors are calculated using the attitude data available via telemetry
and the target values of Delta~V reported by the attitude control system for
each thruster firing event. The model for the atmospheric drag is
significantly simpler than the SRP model, because most of the time it has
close to zero impact on the motion. For the whole duration of the mission the
perigee height was below 1500 km (the nominal height of the atmosphere in the
density model used) only several times.

\begin{table*}
  \centering
  \caption{Description of the RadioAstron dynamic model.}
  \begin{tabular}{l | l}
    Perturbation       & Model \\
    \hline
    Gravity      & \\
    \hline
    Static             & EGM-96 up to 75th degree/order \\
    Third bodies       & JPL DE-405 Sun, Moon and planets \\
    Earth tides        & IERS-2003 conv. \\
    General relativity & IERS-2010 conv. \\
                       & \\
    Surface forces   & \\
    \hline
    Solar radiation  & 3 parameters $\alpha_1$, $\mu_1$ and $\alpha_2$ \\
    Earth radiation  & Applied (constant albedo coeff.) \\
    Atmospheric drag & \makecell[l]{GOST R 25645.166-2004 density model, \\
    cannonball force model with one solve-for parameter}\\
                       & \\
    Desaturation of reaction wheels  \\
    \hline
    Delta V Direction  & Telemetry data from the on-board star sensors and
                         attitude file \\
    Delta V magnitude  & Telemetry data for the duration of thruster firing \\
                       & and a priori thrust model
                         
  \end{tabular}
  \label{tab:dyn_model}
\end{table*}

\section{Tracking and on-board observations}
Several types of observational and telemetry data are used to determine
RadioAstron's orbit. Standard radio tracking consists of range and Doppler
measurements at the C-band. This tracking is performed in the two-way mode
with the 64-m antenna at Bear Lakes and the 70-m antenna at Ussuriysk. Each
tracking session is usually several-hour long, and the observations are
separated by 1--3 days. This time interval is much longer than the typical
separation between successive desaturations of the reaction wheels (several
hours), which makes it difficult to achieve the required accuracy of orbit
determination using this type of data alone.

There is another highly valuable type of radio tracking data, which can be
obtained when the SRT is performing a scientific observation. The RadioAstron
spacecraft does not store any of the radio astronomy data collected by the SRT
and transmits it in real time to Earth using its 1.5-m high-gain antenna
during every experiment.

The downlink carrier can be synchronized either to the highly stable signal of
the on-board H-maser or the uplinked signal of a ground H-maser. The
RadioAstron mission is served by two tracking stations capable of receiving
the scientific data from the satellite and transmitting an uplink reference
signal to it: one at Pushchino (Moscow Region, Russia) and another at Green
Bank (West Virginia, USA). Additionally, these tracking stations perform
Doppler measurements of the downlink signal during each scientific observation
conducted by the satellite. The quality of these measurements is usually much
higher than that of the standard radio tracking at the C-band (see
Table~\ref{tab:radio} below). Since the scientific observations are usually
performed several times a day and last for about an hour, these measurements
provide very valuable data for OD.

A comparison of a posteriori estimated accuracy of radio tracking observations
is shown in Table \ref{tab:radio}. One-way Doppler data provided by the
Pushchino and Green Bank stations are much more accurate than the two-way
Doppler data obtained by the regular tracking stations of Bear Lakes and
Ussuriysk. Relatively large two-way Doppler noise at Bear Lakes and Ussuriysk
is explained by the performance of the C-band telemetry, tracking and control
system used in the project. However, these stations are also capable of
performing range measurements and their two-way Doppler data are free from the
contribution of the a priori unknown frequency offset of the on-board H-maser.

\begin{table}[htbp]
  \centering
  \caption{A posteriori estimates of radio tracking errors based on OD results
    from September 2016 to April 2017. Doppler observations at Bear Lakes and
    Ussuriysk are received over 1~second integration interval, at Pushchino
    and Green Bank --- normal points with 1 minute averaging were employed.}
  \begin{tabular}{l | r | r}
    Tracking station   & \makecell[l]{Range bias \\(RMS), m} & \makecell[l]{Doppler \\noise, mm/s} \\
    \hline
    Bear Lakes, 64-m   & 2.9           & 1.21 \\
    Ussuriysk, 70-m    & 13.4          & 4.78 \\
    Pushchino, 22-m    & N/A           & 0.19 \\
    Green Bank, 140-ft & N/A           & 0.04 \\
  \end{tabular}
  \label{tab:radio}
\end{table}

Radio tracking of RadioAstron is supplemented by optical tracking. Routine
optical angular observations are conducted by a number of observatories,
including the ones of the ISON collaboration, Roscosmos facilities, the MASTER
network and many others. More than 40 different telescopes have observed the
spacecraft so far and provided optical measurements of right ascension and
declination of the spacecraft.

Laser ranging to RadioAstron is also possible and provides distance
measurements with cm-level accuracy. However, such observations are performed
only occasionally for two reasons. First, the retroreflector array is mounted
on the spacecraft in such a way that obtaining laser echoes from it is
possible only when the spacecraft is in a specific orientation (or oriented no
more than approximately 10 degrees away from it). Because of this, it is
almost impossible to perform laser ranging simultaneously with scientific
observations, with the exception of the gravity experiment. Second, the
spacecraft spends most of the time at distances exceeding 100,000 km, which
makes it reachable only to the most powerful laser ranging stations, usually
those capable of ranging to the Moon. So far successful laser ranging
observations have been performed by the following observatories: Grasse
(France), Kavkaz (Russia), Yarragadee, Mt. Stromlo (both in Australia) and
Wettzell (Germany).

A test of applying the Planetary Radio Interferometry and Doppler Experiment
(PRIDE), which is an OD technique based on observing spacecrafts using VLBI,
was conducted during the in-orbit checkout of the RadioAstron mission
\cite{duev2015radioastron}. The technique showed the potential for improving
the accuracy of estimating the spacecraft state vector.  However, PRIDE
operations for RadioAstron would require massive involvement of ground-based
radio telescopes capable of observing at the X- and Ku-bands as well as
significant data processing resources. Such efforts were deemed unnecessary
since other ``traditional'' OD techniques proved sufficient.

Several types of data collected on board and available via telemetry proved
helpful in modeling RadioAstron's dynamics and determining its orbit. The
information on the spacecraft attitude and firings of the attitude control
thrusters is essential for modeling the perturbations due to SRP and Delta V's
that result from momentum dumping of the reaction wheels. Finally, also
available via telemetry are the rotation rates of the reaction wheels, which
implicitly contain information on the dynamics of the spacecraft center of
mass. This information can be extracted in the following way.

Dynamics of the center of mass and attitude dynamics are related to each other
by the action of SRP. The attitude control system of the spacecraft uses
reaction wheels instead of more commonly applied control moment
gyroscopes. This allows us to estimate the perturbing torque as follows. Given
that the spacecraft attitude with respect to an inertial frame is constant on
the time interval between $t_1$ and $t_2$, the change of the angular momentum
can be represented in a body-fixed frame as
\begin{equation}
  \label{eq:rw_torq}
  \sum_{i=1}^8I_i\mathbf{a}_i(\Omega_i(t_2) - \Omega_i(t_1)) = \int_{t_1}^{t_2}\mathbf{M}(t)dt,
\end{equation}
where $I_i$ is the moment of inertia of the $i$-th wheel, $\mathbf{a}_i$ is
its axis of rotation, $\Omega_i(t)$ is its angular velocity, and
$\mathbf{M}(t)$ is the perturbing torque. SRP is usually the only major source
of perturbing torque which is responsible for the change of angular momentum
of the reaction wheels. Therefore, Equation~\eqref{eq:rw_torq} allows one to
obtain the observed value of the SRP torque in a body-fixed frame as long as
the SRP variations on the time interval of $(t_1, t_2)$ are negligible
\begin{equation}
  \label{eq:srp_t_o}
  \mathbf{M}_{SRP,o} = \frac{1}{t_2 - t_1}\sum_{i=1}^8I_i\mathbf{a}_i(\Omega_i(t_2) - \Omega_i(t_1)).
\end{equation}
The corresponding computed value of the SRP torque is given in
Equation~\eqref{eq:srp_t}. The computed value depends on the SRP coefficients
introduced in Section 2, thus the reaction wheel rotation rates contain
valuable information, which otherwise may only be obtained from the orbital
dynamics of RadioAstron.

\section{Orbit determination technique}
According to the dynamic model described above, RadioAstron's orbital motion
is determined by the following vector of parameters
$\mathbf{Q}_d = \{\mathbf{X}(t_0), \alpha_1, \mu_1, \alpha_2,
\Delta\mathbf{v}_1, \ldots, \Delta\mathbf{v}_n\}$, which are, respectively,
the initial state vector of the spacecraft, the three SRP coefficients and the
$n$ vectors of velocity changes due to desaturations of the reaction wheels,
and where $n$ is the number of desaturation events in the orbit determination
time interval.

The kinematic parameters that affect only the computed values of observables
are the session-wise range biases and the frequency offset of the on-board
H-maser, also assumed to be piece-wise constant in a way described in Section
6. The latter affects the one-way Doppler observables obtained by the
Pushchino and Green Bank tracking stations. Standard orbit determination
interval includes at least two full orbits of the spacecraft and usually does
not exceed 30 days.

Orbit determination is performed using the batch least squares estimator. The
observations of the torque and a priori Delta V vectors are assumed to be
independent from each other and from the tracking data. Therefore, the
functional to be minimized can be represented as
\begin{align}
  \label{eq:functional}
  \Phi = &(\bm\Psi_o-\bm\Psi_c)^{\mathsf{T}}\mathbf{P}(\bm\Psi_o-\bm\Psi_c) + \nonumber\\
 + &\sum_{j=1}^{N}(\mathbf{M}_{SRP,o}^{j} - \mathbf{M}_{SRP}^{j})^{\mathsf{T}}\mathbf{P}^{SRP}_j (\mathbf{M}_{SRP,o}^{j} - \mathbf{M}_{SRP}^{j}) + \nonumber\\
  + &\sum_{i=1}^m(\Delta\mathbf{v}^0_i-\Delta\mathbf{v}_i)^{\mathsf{T}}\mathbf{P}_i(\Delta\mathbf{v}^0_i-\Delta\mathbf{v}_i).
\end{align}
The first term in Equation~\eqref{eq:functional} gives the contribution of the
tracking data $\bm\Psi_o$ with corresponding computed values $\bm\Psi_c$ and
inverted covariance estimate of $\bm\Psi_o$ as weight matrix $\mathbf{P}$.

The second term in Equation~\eqref{eq:functional} contains the sum of squares
of residual torques, the observed and computed values of which were described
in the previous section (Equations~\eqref{eq:srp_t_o} and
\eqref{eq:srp_t}).

\begin{figure}[htb]
  \centering
  \includegraphics[width=\linewidth]{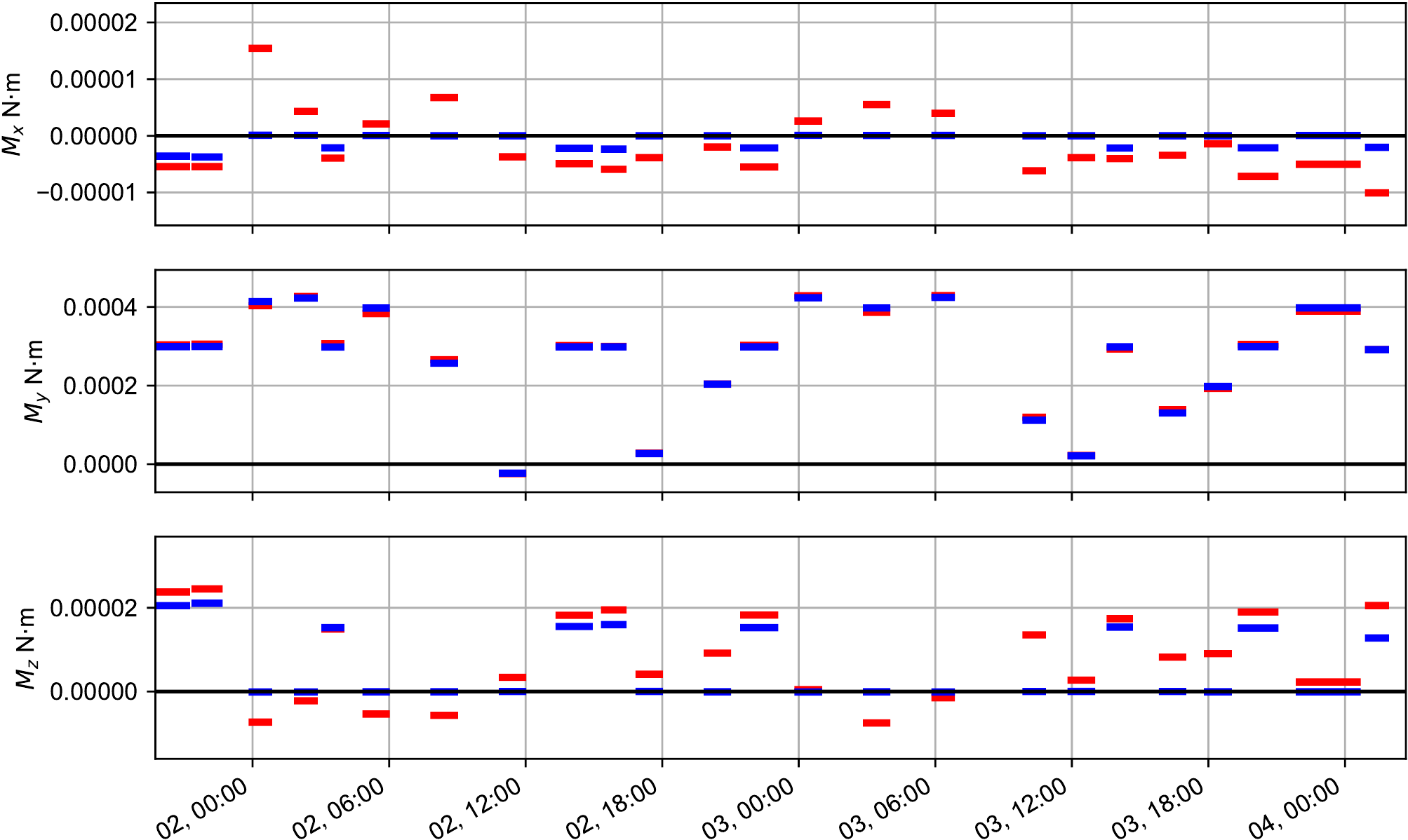}
  \caption{Components of the SRP torque on a 48 hr interval in January 2015:
    observed (red), computed (blue).}
  \label{fig:torq}
\end{figure}

The covariance matrices and their inverses, $\mathbf{P}_{SRP}^j$, can be
estimated using Equation~\eqref{eq:srp_t_o}, given the estimation errors of
the rotation rates of the reaction wheels and the orientations of their axes
are known. In practice, however, the weights have to be adjusted to an
accuracy level of about $2\cdot10^{-6}$ N$\cdot$m, since the perturbing torque
model described above is relatively simple and does not always match the
accuracy of observations. Fig.~\ref{fig:torq} shows the agreement between the
modeled torque and the observational data. It is clear that for the $X$ and
$Z$ components the degree of agreement, in relative terms, is significantly
lower than that for the $Y$ component, but it is the latter that contains the
dominant part of the torque due to SRP.

The last term of Equation~\eqref{eq:functional} takes into account the a
priori information on Delta~V's due to desaturations. For each desaturation
event the value of net $\Delta\mathbf{v}_i^0$ is calculated using information
on every thruster firing that occurred during that event, including its
duration, the fuel pressure and the a priori thrust model. Each of the weight
matrices, $\mathbf{P}_i$, in \eqref{eq:functional} is the inverse of the
respective covariance matrix given by
\begin{align*}
  \mathbf{C} = \sigma_d^2(\mathbf{E} - \mathbf{e}\cdot\mathbf{e}^{\mathsf{T}}) +
  \sigma_m^2\mathbf{e}\cdot\mathbf{e}^{\mathsf{T}}.
\end{align*}
Here, $\mathbf{e}$ is the unit vector of $\Delta\mathbf{v}$ which has the same
direction as X-axis of the spacecraft, $\mathbf{E}$ is a $3\times3$ identity
matrix, $\sigma_m$ is the error in the estimated magnitude of
$\Delta\mathbf{v}$, and $\sigma_d$ is the error in the estimate of
$\Delta\mathbf{v}$ components in the plane orthogonal to $\mathbf{e}$. The
typical value of magnitude error $\sigma_m$ used in the OD is 5\% of
$\Delta v$, the value of the orthogonal component, $\sigma_d$, corresponds to
a 0.25 deg error in direction or $4.36\cdot 10^{-3}$ of $\Delta v$.

In reality the functional is not limited to the one in
Equation~\eqref{eq:functional}. It accounts also for the a priori information
on range biases and the on-board H-maser frequency offset.

\section{ASC correlator overview}

Orbital solutions obtained as a result of the OD procedure outlined above are
used mostly for scientific applications, primarily for correlation of VLBI
data collected by the mission. In this section we review the operations of one
of the correlators capable of processing RadioAstron data, the ASC
Correlator. This FX correlator has been developed by the Astro Space Center of
the Lebedev Physical Institute (ASC LPI) \cite{asc_correlator_2017}
specifically to support the RadioAstron mission. It is used to process the
majority of observations performed by RadioAstron ($\sim$95\%) and thus
provides the largest data set of the so-called residual delays and delay
rates, which we will use in the next section to evaluate the orbit
determination accuracy.

Correlation of space-VLBI data is performed by the ASC Correlator in two
steps, with the second step being necessary due to the orbit determination
uncertainties. In order to decrease the residual delay, delay rate and delay
acceleration, the first step, or pass, is performed in the so-called ``wide''
window mode. The window size along the delay is determined by the number of
spectral channels, while along the fringe rate it is determined by the
integration time. Both parameters are set before the correlation process is
started and can be easily adjusted. After a fringe was found, the second
correlation pass is performed in a smaller window taking into account the
residual delays obtained in the first pass.

RadioAstron's capability to simultaneously observe at two different
wavelengths is of great value for performing the fringe search. In case an
observation was performed at two different wavelengths, the residual values
obtained as a result of the successful fringe search at the longer wavelength
can usually be used to significantly simplify the fringe search at the shorter
one.

A delay model for space-VLBI is naturally more complicated than those used for
ground-based VLBI. Apart from issues related to the orbit, it has to take into
account the fact that the delay rate depends not only on relative velocities
of the telescopes but also on the frequency offset of the uplink signal, which
is used on board as a reference for the on-board scientific equipment (for the
two-way phase-locked loop mode of synchronization), or on the frequency offset
of the on-board H-maser (for the one-way mode). The latter, as it turned out,
could result in delay rates as high as $30-35$ ps/s.

A distinctive feature of the ASC Correlator is its delay model
ORBITA2012. This delay model was developed specifically for the ASC Correlator
and is capable of calculating the delay up to $O(c^{-3})$ terms
\cite{Vlasov2012}, which provides for using smaller correlation window sizes
at the first pass of the correlation process. Moreover, this delay model is
capable of taking into account the delay acceleration.

The post-correlation data reduction for all experiments considered in this
paper was performed using the PIMA package \cite{petrov_pima}. This included
baseline fringe fitting, bandpass and amplitude calibration, and averaging the
data over time and frequency. The primary goal of the post-correlation
processing is to allow for studying the physical properties of the observed
objects. The residual values of group delay, delay rate and delay acceleration
are a byproduct of this processing. The total delay and its derivatives can be
further used for studying the observed objects, e.g. in imaging
experiments. This requires accurate calibration of a number of uncertainties,
e.g. the rates of the telescope clocks, their trajectories (including that of
the SRT), coordinates of the observed sources, propagation media parameters,
etc.

The residual delays and their derivatives, which we used in this analysis,
were obtained as a result of correlation of the observations of the
RadioAstron AGN survey science program and were not used in the OD
procedure. The sources observed in this program are compact enough, so that
the contribution to residual values due to non-point structure of the sources
is negligible.

\section{Orbit determination results}
The orbit determination algorithm outlined in this paper has been implemented
and used by the Keldysh Institute of Applied Mathematics (KIAM) to provide
RadioAstron users with a posteriori orbits required to correlate VLBI data
collected by the mission. The majority of observations performed since 2014
and a significant portion of earlier observations were correlated using orbits
produced with this version of the algorithm. At the early stages of the
mission, i.e. for experiments conducted before 2014, a previous, less
sophisticated version of the OD algorithm was used.

\begin{figure}[htb]
  \centering \includegraphics[width=\linewidth]{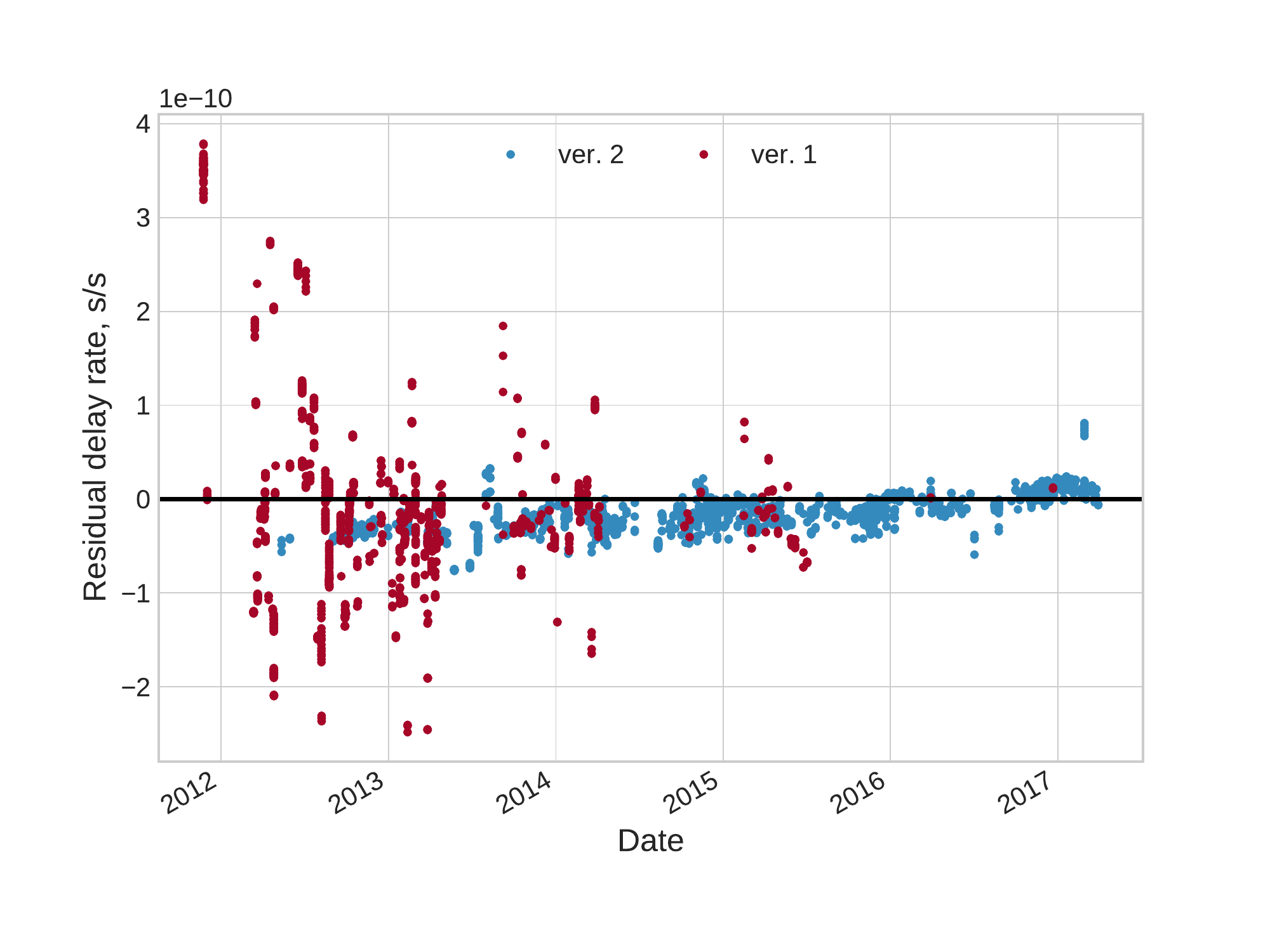}
  \caption{Residual delay rates obtained by the ASC Correlator using two
    versions of the orbit: generic algorithm, version 1 (red dots); new
    algorithm, version 2 (blue dots).}
  \label{fig:rate_all}
\end{figure}

A statistically significant correlation detected between signals recorded by
the RadioAstron space radio telescope and ground-based radio telescopes yields
a set of three residual values, that of delay, delay rate and delay
acceleration. In this paper we focus exclusively on the delay rates for the
following reasons. First, data collected by the space radio telescope are not
time-tagged on board, i.e. synchronization of the on-board H-maser time scale
with that of UTC is possible only indirectly, by means of time-tagging the
data stream flowing from the satellite by the receiving tracking
station. Second, there is a non-zero probability of inconsistent
synchronization of ground radio telescopes with UTC. And finally, as shown
below, there are a number of significant orbit-unrelated factors, which can
contribute to the residual delay rate also affecting the residual delay. All
this makes it difficult to separate the contribution to residual delay due to
errors in the orbit from those due to the involved clocks. On the other hand,
possible contributions to the residual delay rate are much more
predictable. Time synchronization by means of a tracking station does not
affect the delay rate, since any change in the delay it may introduce remains
constant during every ``scan'' (file) of the recorded data. Drifts of
ground-based clocks are usually much smaller than the residual delay rates we
observe and, moreover, when an experiment involves more than one ground radio
telescope, relative drifts of the ground-based clocks can be estimated by
performing fringe searches on ground-only baselines. Finally, the clock drift
of the spacecraft clock can be estimated independently in the process of orbit
determination.

Fig.~\ref{fig:rate_all} shows the residual delay rates obtained by the ASC
Correlator as a result of processing the observations performed by the
RadioAstron mission up to early 2017. Each data point in
Fig.~\ref{fig:rate_all} corresponds to a value of the residual delay rate
determined by the correlator as a result of correlating two data scans,
i.e. units of data typically of $\sim$10 min duration, one of which was
recorded by the RadioAstron space radio telescope and the other by a ground
radio telescope that participated in the experiment. Blue dots denote the
residual delay rates obtained with version 2 orbits, i.e. those determined
with the algorithm outlined in Section 4. Red dots denote the values obtained
with verion 1 orbits, i.e. determined with a generic OD algorithm and a
simpler dynamic model of the spacecraft. The earlier version of the algorithm
uses a simple cannonball model to estimate the SRP and thus ignores its
attitude dependence. Also, the velocity changes due to momentum dumps are not
estimated in the generic algorithm but set to their a priori values,
$\Delta \mathbf{v}_i^0$. For each data point of Fig.~\ref{fig:rate_all} it is
assumed that RadioAstron is the first antenna while the ground radio telescope
is the second one, i.e. the depicted residual delay rates are relative to the
space-based antenna.

All the observations used to obtain the residual delay rates of
Fig.~\ref{fig:rate_all} were conducted in the so-called ``one-way'' mode,
which is characterized by the following two conditions: a) the on-board
scientific equipment is synchronized to the reference signal of the on-board
H-maser; b) the carrier of the downlink signal used to transfer the science
data from the spacecraft to a tracking station is also synchronized to the
reference H-maser signal.

\begin{figure}[htb]
  \centering
  \includegraphics[width=\linewidth]{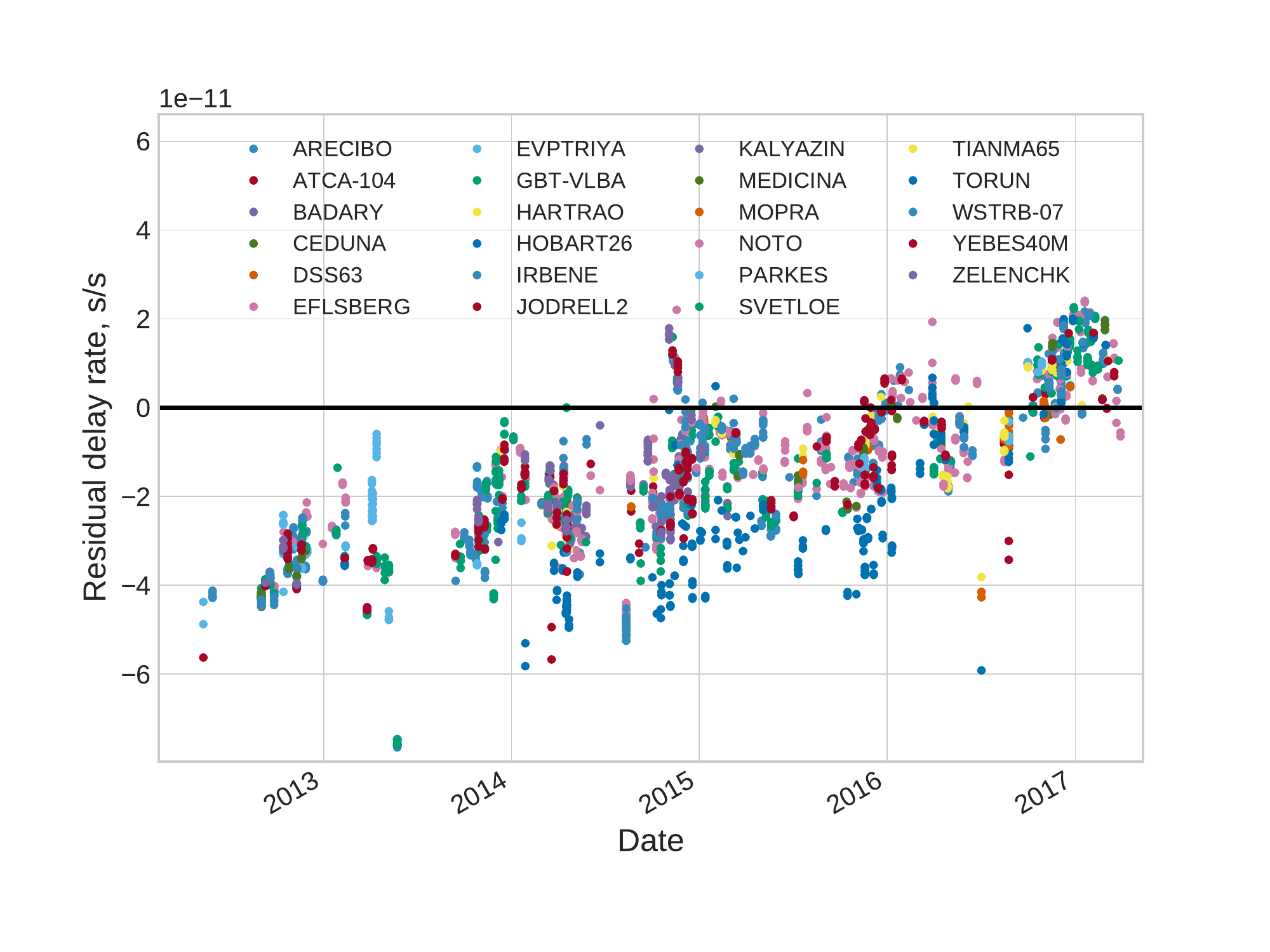}
  \caption{The residual delay rates obtained using version 2 orbits. Color
    indicates the ground radio telescope of the baseline.}
  \label{fig:rrates_v02}
\end{figure}

It is clear from Fig.~\ref{fig:rate_all} that the variation of the residual
delay rates obtained with version 2 of the OD algorithm is substantially
smaller than that of the delay rates obtained with the generic algorithm. The
RMS of the residual delay rates are $2.463\times10^{-11}$ for version 2 and
$1.097\times10^{-10}$ for version 1. It is also clear that the residual delay
rates obtained with version 2 orbits are offset from zero, with the mean value
equal to $1.663\times10^{-11}$ and also reveal a long-term linear trend
(Fig.~\ref{fig:rrates_v02}). The latter is clearly an indication of a
systematic effect which cannot be attributed to an OD error since these
residual values were obtained using a large number of independent orbit
solutions and an even greater number of observations of radio sources
scattered all over the sky and observed at various projected baselines.

Let us consider the impact of the orbit errors on the calculated value of the
group delay reported by the correlator. Denote the spacecraft state vector in
a geocentric inertial reference frame according to the orbital solution by
$\mathbf{X}(t) = (\mathbf{r}_1(t)^{\mathsf{T}},
\mathbf{v}_1(t)^{\mathsf{T}})^{\mathsf{T}}$, the true state vector by
$\overline{\mathbf{X}}(t) = (\overline{\mathbf{r}}_1^{\mathsf{T}}(t),
\overline{\mathbf{v}}_1^{\mathsf{T}}(t))^{\mathsf{T}}$ and the difference
between the two by
$\delta\mathbf{X}(t) = \overline{\mathbf{X}}(t) - \mathbf{X}(t) =
(\delta\mathbf{r}_1(t)^{\mathsf{T}},
\delta\mathbf{v}_1(t)^{\mathsf{T}})^{\mathsf{T}}$. Provided that the
requirements to OD accuracy stated in Section~\ref{intro} are met, the error
in the Geocentric Coordinate Time (TCG) delay rate \cite{Vlasov2012} due to OD
errors can be given by the following equation, which is accurate to at least
$10^{-14}$:
\begin{align}
  \label{eq:rate_corr_tcg}
  \frac{d}{dt}\delta\tau_0 = -\frac{1}{c}\mathbf{k}\cdot\delta\mathbf{v}_1(t_2),
\end{align}
where $\mathbf{k}$ is a unit vector pointing in the direction of the source
(the aberration factor can be omitted provided our accuracy requirements).

The delay understood by the correlator is a difference of proper time
intervals counted by the station clocks from the moment of synchronization to
the moment of signal reception. Thus the estimate for the TCG delay rate given
by Equation~\eqref{eq:rate_corr_tcg} may be insufficiently accurate,
especially in the case when one of the stations is orbiting the
Earth. Assuming the clocks were synchronized at $t_{s}$ the delay should be as
follows
\begin{align}
  \label{eq:delay_proper}
  \tau = \tau_0 + \frac{1}{c^2}\int_{t_s}^{t_2}
  \left(\frac{v_2^2}{2} + U(\mathbf{r}_2)\right)dt
  -\frac{1}{c^2}\int_{t_s}^{t_1}
  \left(\frac{v_1^2}{2} + U(\mathbf{r}_1)\right.\nonumber\\
  \left.+ V(\mathbf{R}_1) - V(\mathbf{R_{\oplus}}) -
  \frac{\partial V(\mathbf{R}_{\oplus})}{\partial\mathbf{R}}\mathbf{r}_1
  \right)dt,
\end{align}
where $\tau_0$ is the TCG delay, $t_1$ and $t_2$ are, respectively, the
moments of signal reception by station 1 and station 2, $U(\mathbf{r})$ is the
Newtonian potential of the Earth at position $\mathbf{r}$, $V$ is the sum of
the Newtonian potentials of all bodies of the Solar System excluding the
Earth, $\mathbf{R}_1$ is the barycentric position of the spacecraft, and
$\mathbf{R}_{\oplus}$ is the barycentric position of the geocenter. The tidal
potential is neglected for the ground-based station (i.e. station 2),
following the recommendations of \cite{iers2010}.

The change of the delay given by Equation~\eqref{eq:delay_proper} due to
errors in the spacecraft position, $\delta\mathbf{r}_1$, and velocity,
$\delta\mathbf{v}_1$, can be represented as
\begin{align}
  \label{eq:delay_proper_corr}
  \delta\tau(\delta\mathbf{r}_1, \delta\mathbf{v}_1) =
  &\delta\tau_0(\delta\mathbf{r}_1, \delta\mathbf{v}_1))\nonumber\\
  &- \frac{1}{c^2}\int_{t_s}^{t_1}\left(\mathbf{v}_1\cdot\delta\mathbf{v}_1
    - U\frac{\mathbf{r}_1\cdot\delta\mathbf{r}_1}{r_1^2}\right)dt,
\end{align}
where the tidal term, which can be approximated by
$\frac{1}{2}\sum_{i=1}^{3}\sum_{j=1}^{3}{\frac{\partial^2V}{\partial
    R_i\partial R_j}r_ir_j}$, has been neglected since for RadioAstron's orbit
and the expected magnitude of OD errors, the contribution of this term to the
delay rate is less than $10^{-19}$. Therefore the correction to the delay rate
can be written as
\begin{align}
  \label{eq:rate_corr_proper}
  \frac{d}{dt}\delta\tau =
  \frac{d}{dt}\delta\tau_0
  - \frac{\mathbf{v}_1\cdot\delta\mathbf{v}_1}{c^2}
  - U\frac{\mathbf{r}_1\cdot\delta\mathbf{r}_1}{c^2r_1^2}.
\end{align}
Provided that OD errors do not exceed the values stated in
Section~\ref{intro}, the last two terms on the right-hand side of
Equation~\eqref{eq:rate_corr_proper} contribute at most $10^{-13}$. Therefore,
if we aim at evaluating the correction to the delay rate with an accuracy of
$10^{-13}$, these two terms can be neglected. Adding to this the correction
due to the on-board H-maser clock drift, $h_1(t)$ and that of the ground-based
station, $h_2(t)$, we arrive at the following equation
\begin{align}
  \label{eq:rate_err}
  \frac{d}{dt}\delta((t_2 - t_s)_p - (t_1 - t_s)_p)
  = \frac{1}{c}\mathbf{k}\cdot\delta\mathbf{v}_1(t_2) - h_1(t_1) + h_2(t_2),
\end{align}
which can be used to investigate the spacecraft velocity error, provided it is
larger than ${\sim}10^{-13}$ but within the limit given in
Section~\ref{intro}, i.e. less than ${\sim}10^{-10}$. The two additional
assumptions needed to justify the use of Equation~\eqref{eq:rate_err} for this
purpose are: a) the position and velocity of the ground-based station (\#2)
are known much better than those of the spacecraft and b) the error of
modeling the contribution of the propagation media to the delay rate is less
than $10^{-13}$. While the first assumption is undoubtedly true, the second
may be violated in rare cases of extreme weather conditions.

The delay model implemented by the ASC Correlator assumes that the spacecraft
and station clocks are ideal, i.e. $h_1(t) \equiv 0$ and $h_2(t) \equiv
0$. Therefore, any instrumental effect affecting the clock drift, if present,
would manifest itself in an extra residual delay rate reported by the
correlator.

If we assume that no other systematic effects are present, each data point
shown in Fig.~\ref{fig:rate_all} corresponds to the right-hand side of
Equation~\eqref{eq:rate_err} evaluated at a certain time using a certain
orbital solution plus an error of the delay rate determination inherent to the
fringe search procedure. Therefore, if we are able to evaluate the clock
drifts of the space and ground-based station clocks, and apply the
corresponding corrections to each data point of Fig. 3, the result will be the
spacecraft velocity errors projected onto the directions to the observed
celestial sources.

We assume that the ground-based station clock drift, $h_2(t)$, is always well
below the level of $10^{-13}$ and thus can be neglected, which is justified by
the results of fringe searches at ground-only baselines. Under this
assumption, any clock drift of the ground station, if present, will be
attributed to the error in the spacecraft velocity. The drift of the on-board
clock, $h_1(t)$, is estimated as part of the OD process in the way outlined
below.

\begin{figure}[htb]
  \centering
  \includegraphics[width=\linewidth]{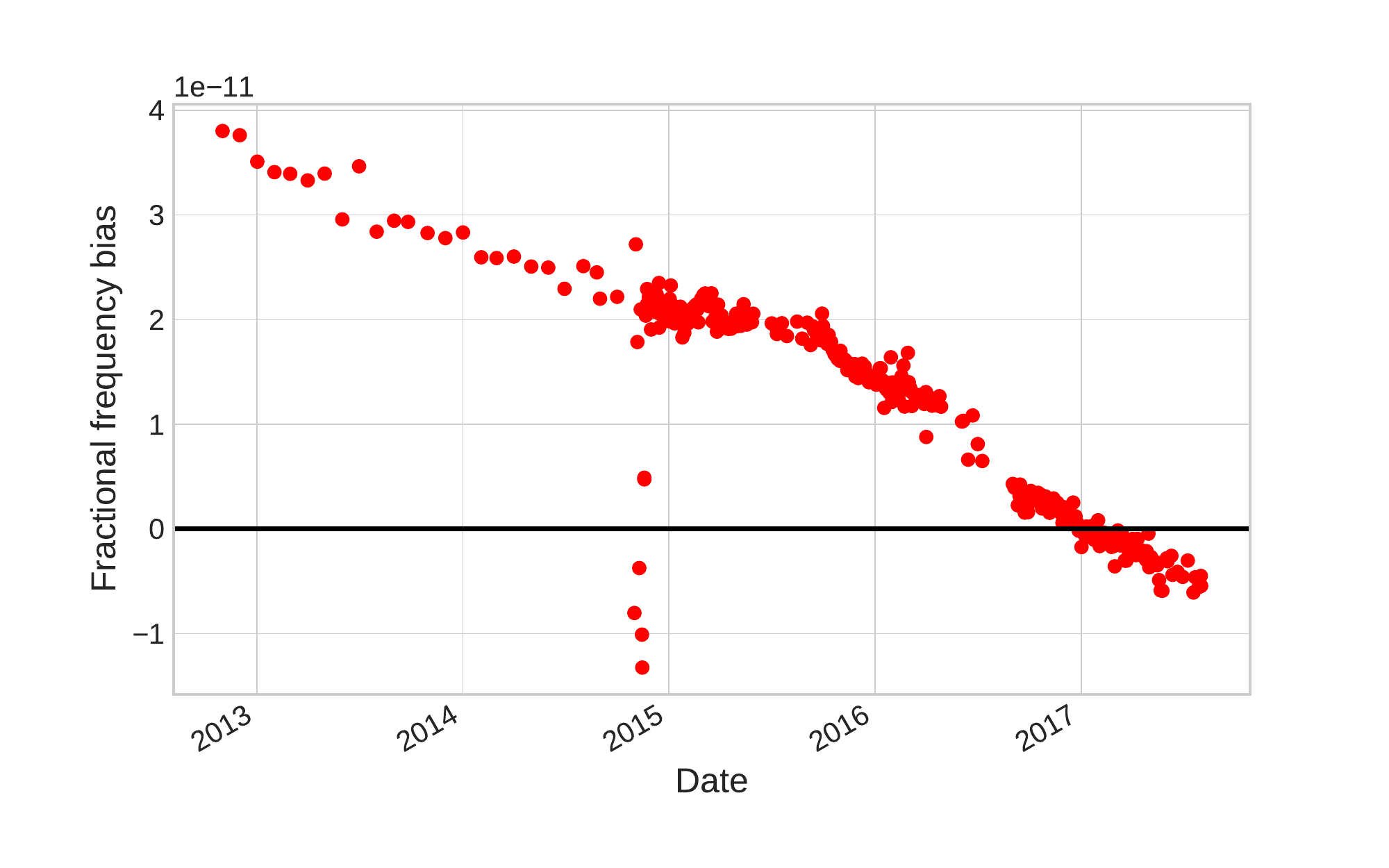}
  \caption{Evolution of the RadioAstron on-board H-maser fractional frequency
    bias. Each value is obtained as part of an orbital solution. The density
    of points before and after October 2014 is different because of a change
    of the frequency bias estimation strategy.}
  \label{fig:df}
\end{figure}

All the observations used for the present analysis were performed, as noted
above, in the so-called one-way mode, i.e. when both the on-board science
equipment and the downlink carrier are locked to the on-board H-maser
reference signal. Therefore, if the on-board H-maser output frequency is
biased, i.e. there is a non-zero clock drift, both the science data, which
eventually results in the residual delay rates used in our analysis, and the
downlink carrier frequency measurements, which are used in OD, will be
affected by this same frequency bias.

In order to properly fit one-way Doppler measurements in the OD process,
frequency bias of the tracking station H-maser and that of the on-board
H-maser have to be taken into account. For each of the two tracking stations
we used a priori values of the frequency biases, which we derived from
long-term series of measured differences between the station's local time,
synchronized to its H-maser, and the GPS time. We assume the on-board H-maser
fractional frequency bias, $\Delta f / f$, to be identical to that of the
downlink carrier, and estimate the latter in the OD process.

The on-board H-maser frequency biases obtained this way are shown in
Fig.~\ref{fig:df}. The bias evolution is almost linear over more than 4 years
of the analyzed data.  In late October 2014 an accident occurred on board
after which the on-board H-maser started experiencing a power surge every time
one of the telemetry and command (T\&C) transponder was turned on. Before that
accident our approach to estimating the on-board H-maser frequency bias was to
treat it as constant over the whole OD interval. After the accident we changed
our strategy and estimated the bias independently for each interval between
the T\&C link transponder on/off switches, which usually coincided with radio
tracking from Bear Lakes or Ussuriysk stations. This change of strategy
explains the different density of data points in Fig.~\ref{fig:df} before and
after October 2014.

\begin{figure}[htb]
  \centering \includegraphics[width=\linewidth]{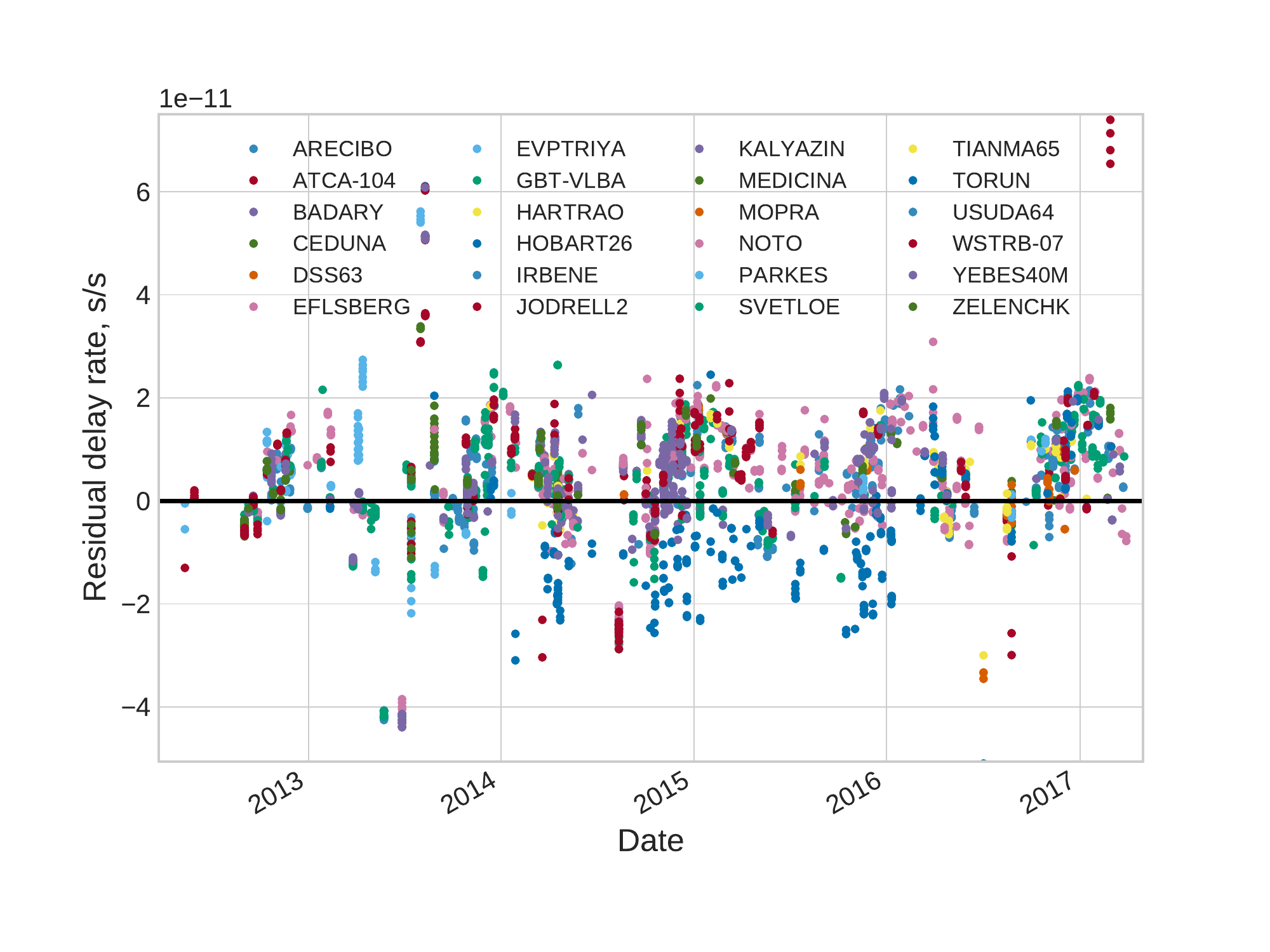}
  \caption{The residual delay rates obtained using version 2 orbits and
    corrected for the on-board clock drift. Color indicates the ground radio
    telescope of the baseline.}
  \label{fig:rrates_corr_clock}
\end{figure}

Using the above assumption that the downlink carrier fractional frequency
bias, $\Delta f/f$, is equal to that of the on-board H-maser, i.e to the
on-board clock drift, $h_1$, we can substract it from the residual delay rates
using Equation~\eqref{eq:rate_err}. The residual delay rates obtained by
applying this correction are shown in Fig.~\ref{fig:rrates_corr_clock}
(cf.~Fig.~\ref{fig:rrates_v02}). With the correction applied the RMS of the
residual delay rates reduces to $1.239\times10^{-11}$ and its mean value
reduces from $1.663\times10^{-11}$ to $3.735\times10^{-12}$.

According to Equation~\eqref{eq:rate_err}, we expect each value of the
residual delay rate corrected for the spacecraft clock drift to be equal to
the corresponding spacecraft velocity error projected onto the direction of
the observed radio source. Therefore, we expect them to be largely
uncorrelated and symmetrically distributed with respect to zero. However, the
values shown in Fig.~\ref{fig:rrates_corr_clock} still reveal significant
systematic effects, i.e. the average value is 1.12~mm/s in terms of velocity
and there are noticeable variations with a period of about 1~year. The
histogram of the data is shown on the left-hand side of
Fig.~\ref{fig:rate_hist}.

\begin{figure*}[htb]
  \centering
  \includegraphics[width=0.49\linewidth]{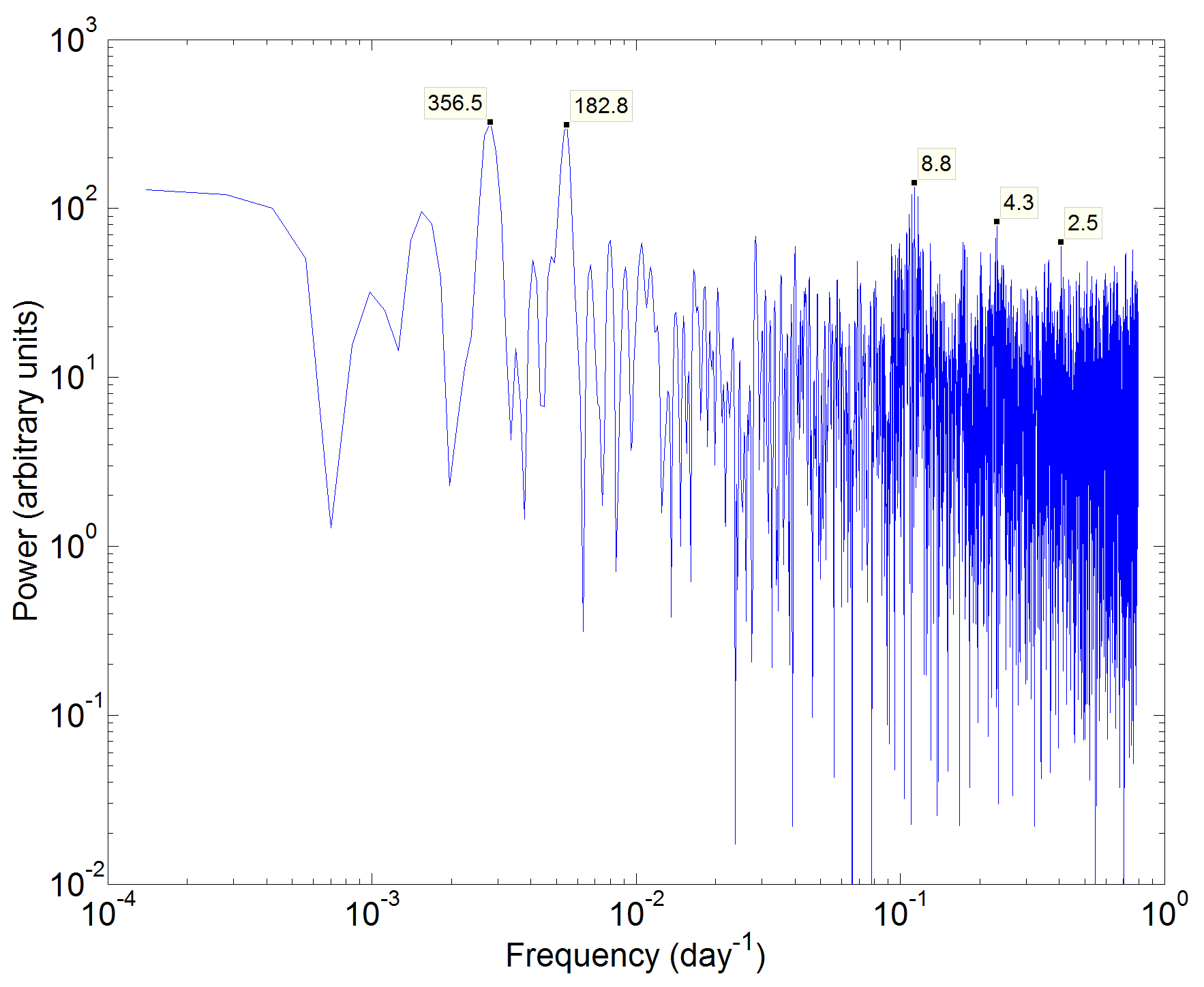}
  \includegraphics[width=0.49\linewidth]{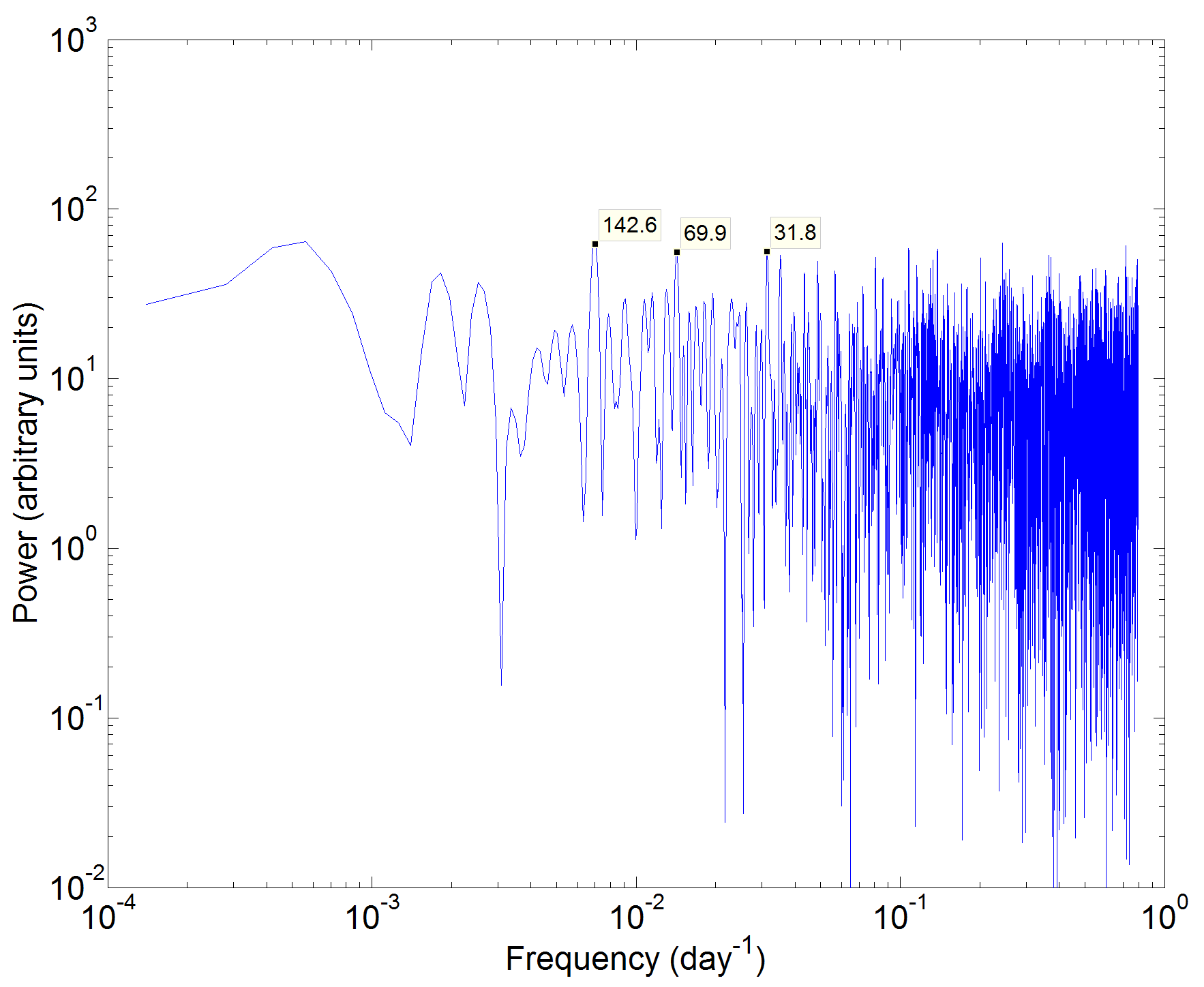}
  \caption{A Lomb-Scargle periodogram of residual delay rates corrected for
    the on-board H-maser drift before (left) and after (right) application of
    the correction of Equation~\eqref{eq:sun_corr}. Significant spikes are
    annotated with the periods (in days) corresponding to their frequencies.}
  \label{fig:pow_spectrum}
\end{figure*}

\begin{figure}[htb]
  \centering
  \includegraphics[width=0.49\linewidth]{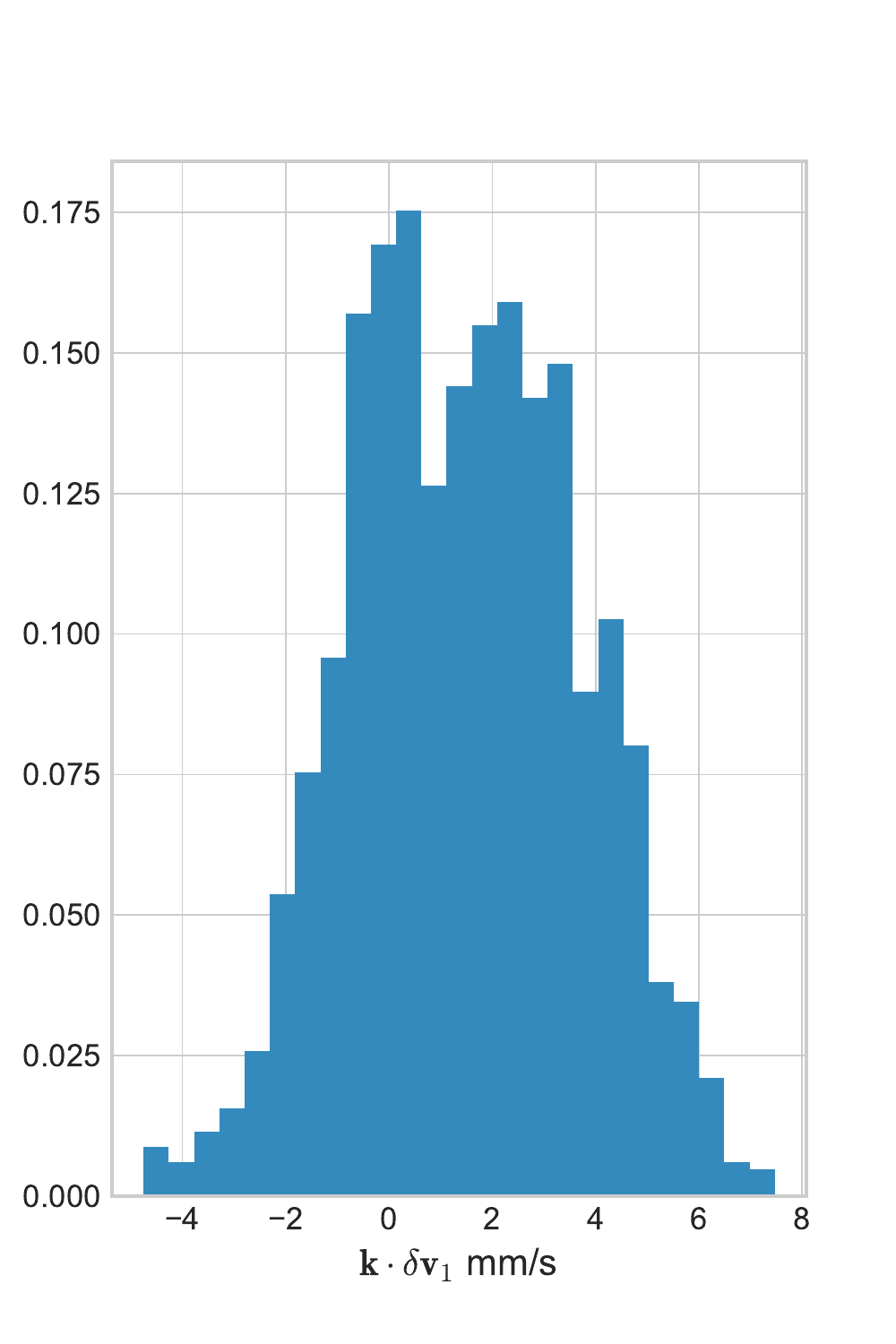}
  \includegraphics[width=0.49\linewidth]{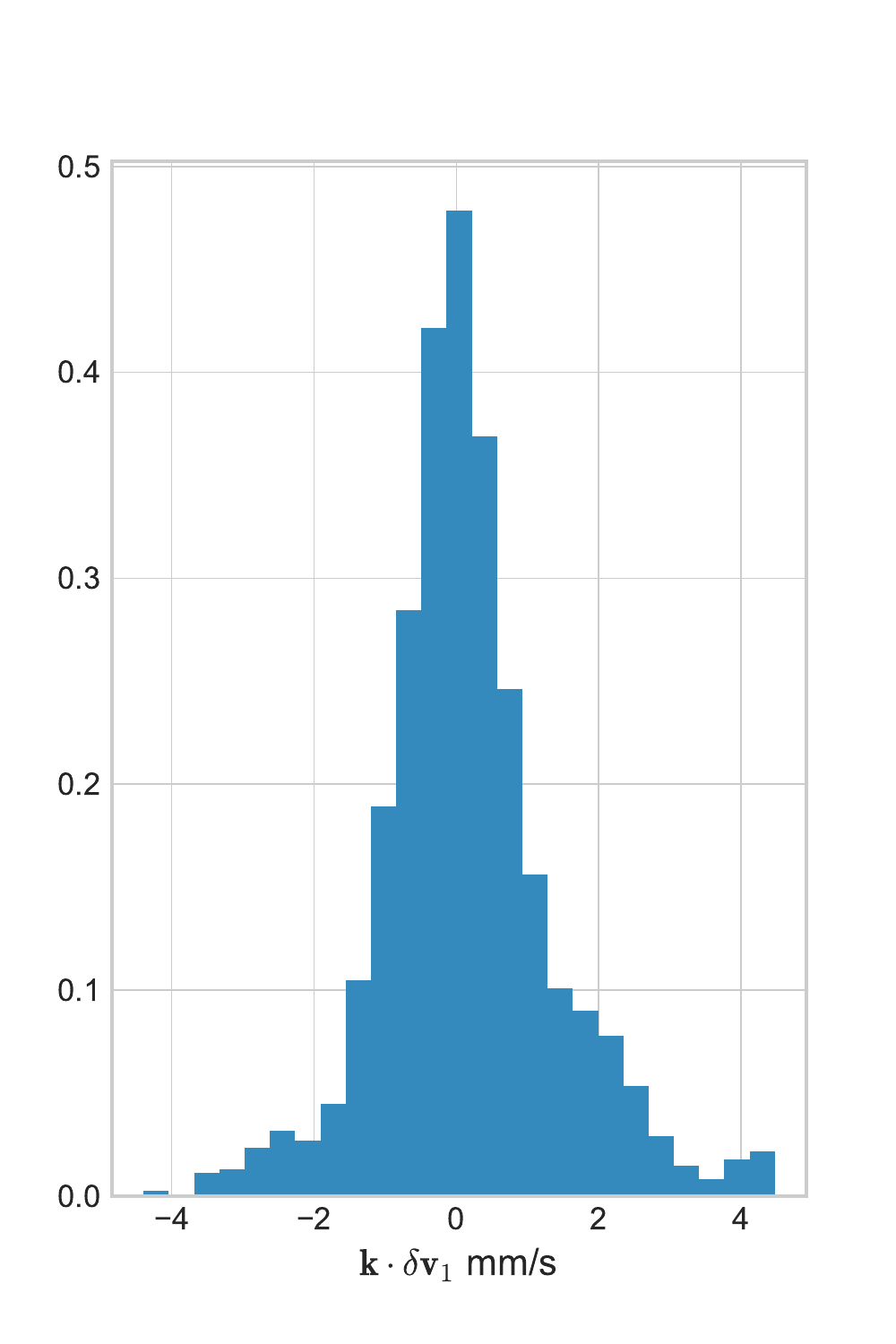}
  \caption{Histograms of the residual delay rates after application of the
    correction for the on-board clock drift. Left: the additional correction
    of Equation~\eqref{eq:sun_corr} is not applied. Right: the correction of
    Equation~\eqref{eq:sun_corr} is applied. Experiments with the Torun ground
    radio telescope are excluded.}
  \label{fig:rate_hist}
\end{figure}

A more detailed analysis of the data shown in Fig.~\ref{fig:rrates_corr_clock}
using the Lomb-Scargle periodogram method reveals that, in addition to the
annual variation, the residual delay rates are modulated with a period of
about 8.8 day (Fig.~\ref{fig:pow_spectrum}), which is close to the average
orbital period.

The periodic variations visible in Fig.~\ref{fig:rrates_corr_clock} and
confirmed on the left panel of Fig.~\ref{fig:pow_spectrum} are present in the
original data and not due to the applied clock correction $h_1(t)$.

In an attempt to understand the origin of these periodicities we tried several
hypotheses. We were able to determine that the periodic patterns in the
residual delay rates can be noticeably reduced by applying the following
correction
\begin{align}
  \label{eq:sun_corr}
  \delta\dot\tau = \frac{1}{c^2}\left(U_{\odot}(\mathbf{R}) - U_{\odot}(\mathbf{R}_{\oplus})\right).
\end{align}
Here, $U_{\odot}$ is the Newtonian potential of the Sun, $\mathbf{R}$ is the
barycentric position of the RadioAstron spacecraft during the experiment and
$\mathbf{R}_{\oplus}$ is the barycentric position of the geocenter.

We are working on developing a full understanding of why the correction of
Equation~\eqref{eq:sun_corr} is required. However, its origin is likely due to
either an inconsistency between the models of orbital motion or the delay
model used by the correlator, or both. This can be concluded from the
following reasoning. Orbital solutions are provided in the geocentric
coordinate system and the Terrestrial Time (TT) scale, as the correlator delay
model expects. In that case the effect of the solar gravitational potential on
the delay is described by the last three terms in the second integral on the
right-hand side of Equation~\eqref{eq:delay_proper}, with the potential $V$
replaced by $U_{\odot}$. For the delay rate it differs from the correction in
Equation~\eqref{eq:sun_corr} by the third term,
${\partial U_{\odot}(\mathbf{R}_{\oplus})}/{\partial\mathbf{R}}\cdot
\mathbf{r}_1/c^2$, which is approximately equal to the expression in
Equation~\eqref{eq:sun_corr} and makes the tidal effect of the solar
gravitational potential almost negligible in the case of
RadioAstron. Therefore, the correction in the form of
Equation~\eqref{eq:sun_corr} cannot be due to an unaccounted effect of the
solar gravitational potential but can only arise due to an inconsistency in
the computation of delays using the provided orbit solutions.

\begin{figure}[htb]
  \centering
  \includegraphics[width=\linewidth]{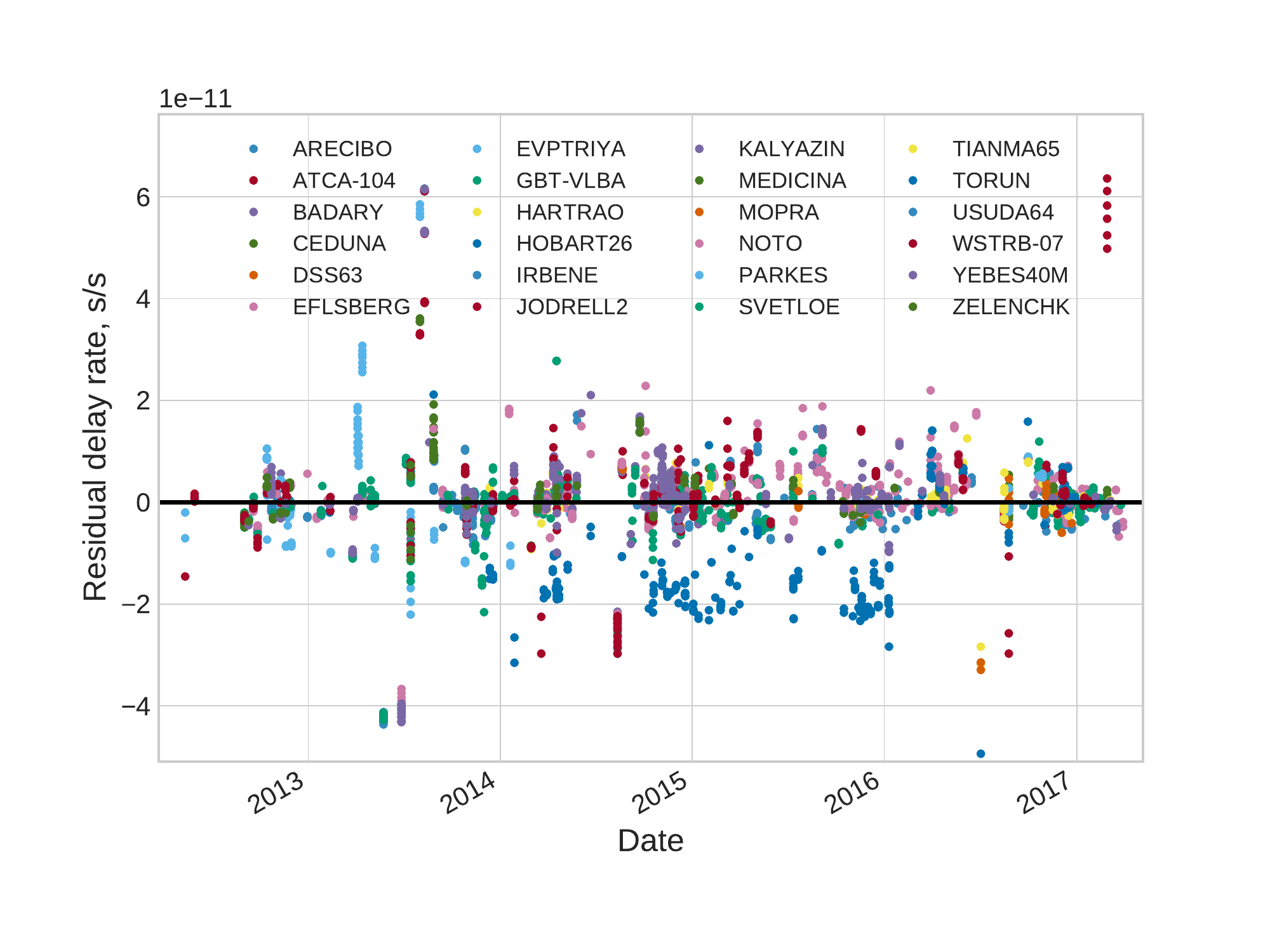}
  \caption{The residual delay rates obtained using version 2 orbits corrected
    for the on-board clock drift and for the correction of
    Equation~\eqref{eq:sun_corr}. Color indicates the ground radio telescope
    of the baseline.}
  \label{fig:rates_final}
\end{figure}

The residual delay rates after application of the correction of
Equation~\eqref{eq:sun_corr} are shown in Fig.~\ref{fig:rates_final}. In
addition to overall reduction of delay rate variation, it is now clear that
part of the data points, i.e. those that correspond to RadioAstron--Torun
baselines, have an appreciable bias of about $-1.9\times10^{-11}$, which can
reasonably be attributed to the clock drift of this ground radio telescope.

Excluding the Torun data, which constitute 8.2\% of the dataset, we arrive at
the mean value of the residual delay rates of just $4.387\times10^{-13}$ and
its RMS of $9.573\times10^{-12}$. The distribution of the data also becomes
more symmetric (Fig.~\ref{fig:rate_hist}).

In order to investigate the statistical properties of the residual delay rates
shown in Fig.~\ref{fig:rates_final} further, we exclude the data of
experiments conducted in the summer periods, i.e. between the 1st of June and
the 1st of September. These experiments constitute 10.4\% of data, but their
RMS, equal to $2.307\times10^{-11}$, is more than 5 times larger than that of
the rest of the data. Degrade orbit determination accuracy is definitely one
of the main reasons why the summer experiments exhibit larger residual delay
rates. The number of scientific observations drops significantly during
summer, mainly due to constrains on the spacecraft attitude with respect to
the Sun, which reduce the visibility of radio sources of interest. The lack of
experiments reduces the amount of Doppler data from the tracking stations
which, in turn, affects the orbit accuracy.

After excluding the summer experiments and filtering out outliers, which, for
the 5$\sigma$ threshold, constitute 1.15\% of the original data, we obtain a
dataset which contains 88.4\% of the initial number of residual delay
rates. For this subset of the data the mean residual delay rate is
$4.428\times10^{-13}$ and the RMS is $4.255\times10^{-12}$. This corresponds
to the standard deviation of the projected spacecraft velocity error,
$\mathbf{k}\cdot\delta\mathbf{v}_1$, of~1.275~mm/s.

To summarize, we obtained a dataset of the spacecraft velocity error estimates
projected onto the directions of the observed sources,
$\mathbf{k} \cdot \delta \mathbf{v}_1$. We assumed that clock drifts of
ground-based VLBI stations that provided the ground legs of the baselines are
negligible, the on-board clock drift is equal to the estimated value of the
downlink carrier fractional frequency offset, and we applied the correction of
Equation~\eqref{eq:sun_corr} to remove systematic patterns in the residual
delay rates due to suspected inconsistencies between the orbit and correlator
delay models. Now, in order to obtain information on the spacecraft velocity
error itself, $\delta \mathbf{v}_1$, and not only on its projection,
$\mathbf{k}\cdot\delta \mathbf{v}_1$, we represent the residual delay rates as
a function of the angle, $\alpha$, between the direction to the observed
source, $\mathbf{k}$, and the direction to the tracking station that was
receiving data from the spacecraft.

\begin{figure}[htb]
  \centering \includegraphics[width=\linewidth]{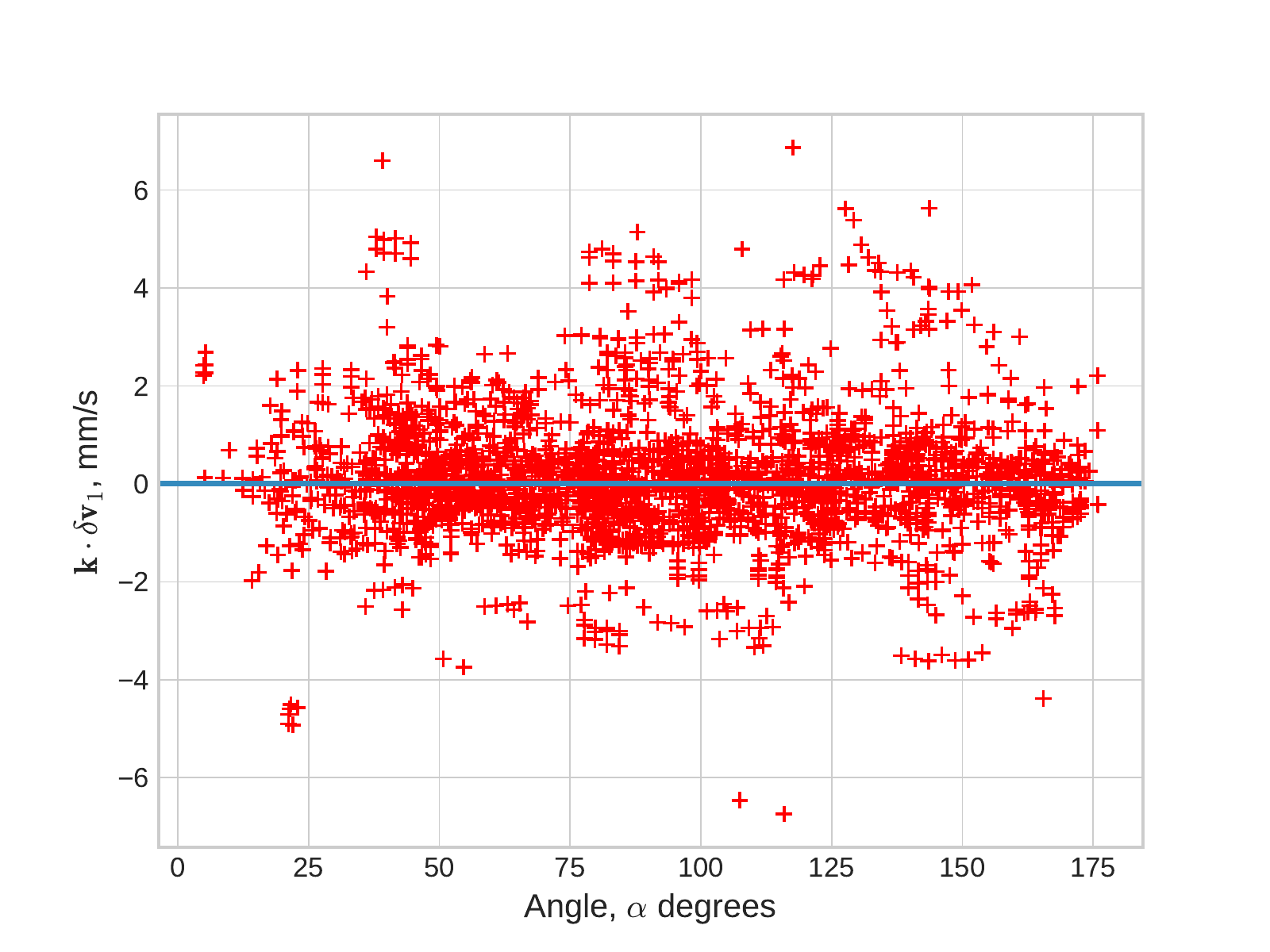}
  \caption{Dependence of the residual delay rates on the angle between the
    direction to observed source and the direction from the spacecraft to the
    tracking station which was used to receive the data of that observation}
  \label{fig:rates_vs_angle}
\end{figure}

Since every science observation is accompanied by simultaneous Doppler
measurements performed by the tracking station used to receive the science
data, the largest errors are to be expected in the velocity components that
belong to the plane which is normal to the tracking station
direction. However, contrary to this expectation,
Fig.~\ref{fig:rates_vs_angle} does not reveal any significant dependence of
the projected velocity error, $\mathbf{k}\cdot\delta \mathbf{v}_1$, on
$\alpha$. A possible reason for this is that the corrected residual delay
rate, and thus $\mathbf{k}\cdot\delta\mathbf{v}_1$, contains errors due to
inaccurate estimation of the on-board clock drift, $h_1(t)$, and the
assumption on the ground station clocks that $h_2(t)\equiv 0$, both of which
do not depend on the source direction, $\mathbf{k}$, and may provide
significant contributions to the residual delay rates.

We can estimate the velocity error along the ``worst'' directions using the
dependence of the projected velocity error,
$\mathbf{k}\cdot\delta \mathbf{v}_1$, on $\alpha$. For the experiments, in
which $75^{\circ} \leq \alpha \leq 105^{\circ}$, the RMS of
$\mathbf{k} \cdot \delta\mathbf{v}_1$ is equal to 1.359 mm/s. Assuming the
spacecraft velocity is determined not worse along other directions the error
ellipsoid of $\delta\mathbf{v}_1$ is confined by the one corresponding to the
covariance matrix $diag(1.85\ mm^2/s^2, 1.85\ mm^2/s^2, 1.85\ mm^2/s^2)$.

In order to obtain an independent estimate of the velocity error, we compared
55 independent adjacent orbit solutions from 2014 to 2018 for discontinuity in
velocity. If the first solution was obtained for, e.g., the time interval of
$(t_1, t_2)$ and the second one for $(t_2, t_3)$, the spacecraft state vector
is compared at $t_2$. The time of comparison, $t_2$, is thus either the
beginning or the end of an OD interval. Unlike the case of the residual delay
rates, these times do not correspond to any tracking activity and, therefore,
the state vectors evaluated at these moments contain additional prediction
errors.

The 3D RMS of the 54 velocity differences computed this way is 4.38~mm/s. Just
like the residual delay rates, these velocity differences exhibit significant
increase in magnitude in the summer months due to the decrease of the amount
of tracking data. If we exclude the summer data, we arrive at a dataset of 42
points and 3D RMS of the velocity differences of 3.52~mm/s.

However, we should take into account that such differences contain errors of
two independent orbit solutions, therefore, their variance is as sum of
variances of two orbit solutions. Thus, we can conclude that the 3D RMS of the
velocity error of a single solution is 3.10~mm/s for the full dataset and
2.53~mm/s for its summer-free subset.

The RMS values of the spacecraft velocity error provided by the discontinuity
analysis and those obtained using the residual delay rates are
similar. However, since the former contain, as noted above, additional
prediction errors, we expected them to be higher. This may be an indication of
the fact that even after the corrections to the residual delay rates have been
applied, clock-related errors still give a significant contribution to these
data, resulting in overestimated values of velocity errors.

\section{Discussion}
We obtained two independent estimates of the velocity errors, the one using
the comparison of adjacent independent orbits and the other via the residual
delay rate analysis. The obtained results agree very well. However, in order
to obtain the velocity error estimates using the residual delay rate data, we
had to apply two non-trivial corrections to the data.

The first correction is to take into account an instrumental effect of the
spacecraft clock drift. The significance of the on-board H-maser clock drift
is evident from the combined analysis of two-way range and Doppler tracking
data together with the one-way Doppler measurements of the downlink carrier
synchronized to the on-board H-maser. Since the ASC Correlator assumes the
clocks to be ideal, i.e. any non-zero clock drift goes directly into the
residual delay rate, the necessity to correct for the on-board clock drift is
well justified. This correction, however, introduces additional errors into
the residual delay rate data due to the way it is estimated in the process of
OD.

The necessity of the second correction, that of Equation~\eqref{eq:sun_corr},
is also clear as it significantly improves the residual delay rates by
cleaning them from artificial periodic patterns. However, its nature is
currently not as well understood. We suspect the necessity of this correction
is due to an inconsistency between the models of orbital motion or the delay
model used by the correlator, or both. Investigation of the physical cause
behind this corrections is beyond the scope of the present work and would
require a deeper insight in the experimental data accumulated by the mission.

The residual delay rate data provide a unique means of estimating velocity
errors of a space-VLBI spacecraft. If the mission performs observations of
radio sources distributed all over the sky, the method effectively provides
estimates of all components of the spacecraft velocity error, as shown in
Fig.~\ref{fig:rates_vs_angle}.

The OD accuracy evaluation of the only previous space-VLBI mission of
VSOP/HALCA, given by \cite{halca_nav}, did not include an analysis of the
residual delay rate data. However, the overall OD accuracy of the HALCA
spacecraft, obtained with traditional methods, was better than that of
RadioAstron. This can be easily understood as the HALCA spacecraft was tracked
by a larger number of tracking stations, each possibly with better
instrumental stability than those involved in RadioAstron.

\section{Conclusion}
We outlined our approach to orbit determination of the RadioAstron
spacecraft. The method includes an in-house developed SRP model and an
algorithm to take into account the accumulated angular momentum of reaction
wheels to improve our knowledge of the dynamics of the spacecraft center of
mass. We tested the performance of this OD method using the unique ``tracking
data'' available only for a space-VLBI spacecraft, i.e. the residual delay
rates, which in our case were obtained by the ASC Correlator of the
RadioAstron mission as a result of processing the data of VLBI observations of
celestial radio sources performed by the RadioAstron spacecraft together with
ground radio telescopes. This analysis allowed us to conclude that the OD
method we developed provides up to 11 times more accurate orbital solutions in
terms of velocity and residual delay rates as compared to a generic OD
algorithm.

Using the residual delay rate data we have obtained an estimate of the
standard deviation for every component of the spacecraft velocity error. The
residual delay rates obtained in non-summer observations, which constitute
88.4\% of all data, exhibit a standard deviation of the spacecraft velocity
error of 1.4 mm/s or less for each velocity component. This result is
consistent with an independent a posteriori estimate of the spacecraft
velocity error obtained using the analysis of velocity differences computed at
the boundaries of adjacent orbital solutions.

\section{Acknowledgements}
The RadioAstron project is led by the Astro Space Center of the Lebedev
Physical Institute of the Russian Academy of Sciences and the Lavochkin
Scientific and Production Association under a contract with the Russian
Federal Space Agency, in collaboration with partner organizations in Russia
and other countries including Keldysh Institute of Applied Mathematics of the
Russian Academy of Sciences. 
Partly based on observations performed with radio telescopes of IAA RAS
(Federal State Budget Scientific Organization Institue of Applied Astronomy of
Russian Academy of Sciences).
Partly based on the Evpatoria RT-70 radio telescope (Ukraine) observations
carried out by the Institute of Radio Astronomy of the National Academy of
Sciences of Ukraine under a contract with the State Space Agency of Ukraine
and by the National Space Facilities Control and Test Center with technical
support by Astro Space Center of Lebedev Physical Institute, Russian Academy
of Sciences.
Partly based on observations with the 100-m telescope of the MPIfR
(Max-Planck-Institute for Radio Astronomy) at Effelsberg.
Partly based on observations with the Medicina (Noto) telescope operated by
INAF - Istituto di Radioastronomia.
The National Radio Astronomy Observatory is a facility of the National Science
Foundation operated under cooperative agreement by Associated Universities,
Inc.
The Green Bank Observatory is a facility of the National Science Foundation
operated under cooperative agreement by Associated Universities, Inc.
The Arecibo Observatory is operated by SRI International under a cooperative
agreement with the National Science Foundation (AST-1100968), and in alliance
with Ana G. Mendez-Universidad Metropolitana, and the Universities Space
Research Association.
The Australia Telescope Compact Array (Parkes radio telescope / Mopra radio
telescope / Long Baseline Array) is part of the Australia Telescope National
Facility which is funded by the Commonwealth of Australia for operation as a
National Facility managed by CSIRO.
This work is based in part on observations carried out using the 32-meter
radio telescope operated by Torun Centre for Astronomy of Nicolaus Copernicus
University in Torun (Poland) and supported by the Polish Ministry of Science
and Higher Education SpUB grant.
Results of optical positioning measurements of the Spektr-R spacecraft by the
global MASTER Robotic Net \cite{lipunov2010}, ISON collaboration, and Kourovka
observatory were used for spacecraft orbit determination in addition to
mission facilities.

\bibliographystyle{elsarticle-harv}
\bibliography{ref3}

\end{document}